\documentclass[twocolumn,twoside]{IEEEtran}

\def\currentdate{Sep.~16, 2016}
\def\revdate{Feb.~8, 2017}
\def\accepteddate{Apr.~2, 2017}

\usepackage{mathpazo}
\usepackage{times}
\usepackage{color}
\usepackage{amsmath}
\usepackage{amsfonts}
\usepackage{latexsym}
\usepackage{amssymb}
\usepackage{cite}

\usepackage{upref}
\usepackage{theorem}
\usepackage{graphicx}
\usepackage{psfrag}
\usepackage{subfigure}

\theoremstyle{plain}
\theorembodyfont{\normalfont\slshape}

\newtheorem{thm}{Theorem$\!$}
\newenvironment{theorem}
{\begin{thm}\hspace*{-1ex}{\bf.}}{\end{thm}}

\newtheorem{lem}[thm]{Lemma$\!$}
\newenvironment{lemma}{\begin{lem}\hspace*{-1ex}{\bf.}}{\end{lem}}

\newtheorem{prop}[thm]{Proposition$\!$}

\newtheorem{cor}[thm]{Corollary$\!$}

\newtheorem{defn}[thm]{Definition$\!$}
\newenvironment{definition}{\begin{defn}\hspace*{-1ex}{\bf.}}{\end{defn}}

\newtheorem{xmpl}[thm]{Example$\!$}

\newtheorem{cnstr}{Construction$\!$}

\newcounter{enumrom}
\renewcommand{\theenumrom}{(\roman{enumrom})}

\makeatletter
\renewcommand{\@endtheorem}{\endtrivlist}
\makeatother

\makeatletter
\renewcommand{\thefigure}{{\@arabic\c@figure}}
\renewcommand{\fnum@figure}{{\bf Figure\,\thefigure}}
\makeatother

\newcommand{\mathset}[1]{\left\{#1\right\}}
\newcommand{\abs}[1]{\left|#1\right|}
\newcommand{\bigabs}[1]{\big|#1\big|}

\newcommand{\floorenv}[1]{\left\lfloor #1 \right\rfloor}
\newcommand{\parenv}[1]{\left( #1 \right)}

\newcommand{\be}[1]{\begin{equation}\label{#1}}
\newcommand{\ee}{\end{equation}}

\renewcommand{\leq}{\leqslant}

\renewcommand{\geq}{\geqslant}

\renewcommand{\Bbb}{\mathbb}

\newcommand{\Cref}[1]{Co\-ro\-lla\-ry\,\ref{#1}}

\renewcommand{\Bbb}{\mathbb}

\newcommand{\R}{{\Bbb R}}

\newcommand{\bu}{\Phi}
\newcommand{\bl}{\varphi}
\DeclareMathOperator{\per}{per}

\DeclareMathOperator{\gap}{Gap}
\newcommand{\that}{\hat{t}}
\newcommand{\ghat}{\hat{g}}

\newcommand{\eg}{\emph{e.g.}}

\newcommand{\ie}{\emph{i.e.}}
\newcommand{\etc}{\emph{etc.}}

\DeclareMathOperator{\id}{Id}
\DeclareMathOperator{\diag}{diag}

\newcommand{\defeq}{\triangleq}

\newcommand*{\defend}{\hfill\ensuremath{\Diamond}}%
\newcommand*{\theoremend}{\hfill\ensuremath{\Box}}%

\outer\def\proclaim #1. #2\par{\medbreak
 \noindent{\bf#1.\enspace}{\sl#2\par}%
 \ifdim\lastskip<\medskipamount \removelastskip\penalty55\medskip\fi}

\begin{document}

\title{\textbf{Improved Lower Bounds on the Size of Balls 
               over Permutations with the Infinity Metric}}

\author{\large
  Moshe~Schwartz,~\IEEEmembership{Senior Member,~IEEE},
  and
  Pascal~O.~Vontobel,~\IEEEmembership{Senior Member,~IEEE}
\thanks{The material in this paper was presented in part at the
        IEEE International Symposium on Information
        Theory (ISIT 2015), Hong Kong, China SAR, 
        June 2015~\cite{SchVon15}.}%
\thanks{Moshe Schwartz is with the 
        Department of Electrical and Computer Engineering, 
        Ben-Gurion University of the Negev,
        Beer Sheva 8410501, Israel
        (e-mail: schwartz@ee.bgu.ac.il).}%
\thanks{Pascal O.~Vontobel is with the 
        Department of Information Engineering, 
        The Chinese University of Hong Kong,
        Shatin, N.T., Hong Kong,
        (e-mail: pascal.vontobel@ieee.org).}%
\thanks{This work was supported in part by the 
        Israel Science Foundation (ISF) grant No.~130/14.} 
\thanks{Submitted to IEEE Transactions on Information Theory, 
        \currentdate. Revised on \revdate. Accepted on \accepteddate.}
}

\maketitle

\begin{abstract}
  We study the size (or volume) of balls in the metric space of
  permutations, $S_n$, under the infinity metric. We focus on the
  regime of balls with radius $r = \rho \cdot (n\!-\!1)$, $\rho \in
  [0,1]$, \ie, a radius that is a constant fraction of the maximum
  possible distance. We provide new lower bounds on the size of such
  balls. These new lower bounds reduce the asymptotic gap to the known
  upper bounds to at most $0.029$ bits per symbol. Additionally, they
  imply an improved ball-packing bound for error-correcting codes, and
  an improved upper bound on the size of optimal covering codes.
\end{abstract}

\begin{IEEEkeywords}
  Asymptotic gap,
  infinity metric,
  permanent,
  permutation,
  rank modulation,
  Sinkhorn theorem.
\end{IEEEkeywords}

\section{Introduction}
\label{sec:intro}

\IEEEPARstart{G}{iven} a metric space $(M,d)$, perhaps one of the most basic
constructs is that of a ball
\begin{align*}
  B_r(x)
    &\defeq
       \mathset{ x'\in M ~|~ d(x,x')\leq r },
\end{align*}
where $x\in M$ is the ball's center and $r$ is the ball's
radius. Since many coding-theoretic problems may be viewed as the
study of packing or covering of a metric space by balls, properties of
balls and their parameters have been studied extensively in a wide
range of metrics \cite{MacSlo78,CohHonLitLob97,ConSlo88}.

An important feature of a ball is its size (or volume), \ie, the
number of points in the ball. It is an important component in many
bounds on code parameters, most notably, the ball-packing bound and
the Gilbert--Varshamov bound~\cite{MacSlo78}. Thus, the exact size,
the asymptotic size, or bounds on the size of balls in various metrics
are of interest.

Lately, metric spaces over permutations have received increased
attention. This is motivated, in particular, by the recent application of rank
modulation to non-volatile memory systems~\cite{JiaMatSchBru09}: in such
applications, the charge levels of memory cells are compared against each
other, and a permutation is induced by the relative ranking of the cells'
charge levels. For designing error-correcting codes or covering codes over the
space of permutations, one needs to choose a suitable metric and so several
metrics have been studied for the space of permutations, including Hamming's
metric
\cite{Bla74,DezFra77,DezVan78,BlaCohDez79,CamWan05,KeeKu06,Qui06,Bai09,Cam10},
Kendall's $\tau$-metric
\cite{ChaKur69,JiaSchBru10,BarMaz10,MazBarZem13,WanMazWor15,BuzEtz14,BuzYaaEtz14,ZhoSchJiaBru15},
and Ulam's metric~\cite{FarSkaMil13,GolLemRieSka15}.

This paper focuses on the infinity metric (whose formal definition
will follow in the next section), which is motivated by applications
to rank modulation in some non-volatile memory systems (\eg, flash
memory or phase-change memory systems). Recall that in a flash memory
system, each cell has a charge level which can be changed or read out
by a suitable circuit. The idea behind rank modulation is that
information is encoded in terms of the rank of the charge values. More
precisely, consider $n$ flash cells, each with some charge level. Let
$f_1$ be the rank of the charge level of the first cell, $f_2$ be the
rank of the charge level of the second flash cell, \etc\ The
information encoded in these $n$ flash cells corresponds to the
permutation $(f_1,f_2,\dots,f_n)$.  However, charge levels may be set
imprecisely during the storing process, they may change over time due
to physical processes, or they may be read out imprecisely due to
noise, thereby giving rise to a distorted permutation
$(f'_1,f'_2,\dots,f'_n)$. In a typical setting, we may have some bound
on the charge-level distortion, which translates to a bound on
$d\defeq\max_{i}\abs{f_i-f'_i}$, \ie, each cell may change its rank by
a limited amount. We say $d$ is the distance between the two
permutations under the infinity metric. With a metric space in hand we
can combat such distortion by designing suitable error-correcting
codes.

More generally, spaces of permutations with the infinity metric have been
used for
error-correction~\cite{TamSch10,ShiTsa10,KloLinTsaTze10,ZhoSchJiaBru15}, code
relabeling~\cite{TamSch12}, anticodes~\cite{SchTam11}, covering codes
\cite{WanMazWor15,FarSchBru16a}, and snake-in-the-box codes
\cite{YehSch12b,YehSch16}. It is therefore surprising that the asymptotic size
of a ball in this metric space is (to the best of our knowledge) unknown, and
a considerable gap exists between the known lower and upper bounds.

The goal of this paper is to reduce the gap between the lower and
upper bounds on the asymptotic size of balls in the space of
permutations with the infinity metric. To that end, we exploit a
well-known connection between the size of the aforementioned balls,
and permanents of binary Toeplitz matrices. We carefully employ lower
bounds on permanents of non-negative matrices to obtain the desired
results. (One of these bounds is well known, one is somewhat recent.)

The paper is organized as follows. In Section~\ref{sec:notation} we
present notations and definitions. Result-wise, the main section is
Section~\ref{sec:results:1}, where we collect not only known, but also
our new results on the asymptotic gap between upper and lower
bounds. Whereas in Section~\ref{sec:prelim} we discuss how the known
results in Section~\ref{sec:results:1} are obtained, we devote
Section~\ref{sec:newbound} to the presentation of the new lower bounds
that lead to the new results in Section~\ref{sec:results:1}. We
conclude the paper in Section~\ref{sec:conc}.

\section{Notation}
\label{sec:notation}

For the rest of this paper, $n$ will denote a positive integer. With
this, we define $[n] \defeq \mathset{1,2,\dots,n}$ and let $S_n$ be
the set of all permutations over $[n]$. The identity permutation in
$S_n$ is denoted by $\id_n$. Additionally, the composition of any two
permutations $f,g\in S_n$ is denoted by $fg$ and represents the
mapping $i\mapsto f(g(i))$.

For any $f,g\in S_n$, the infinity metric (or infinity
distance) between them, denoted $d_\infty(f,g)$, is defined as
\begin{align*}
  d_\infty(f,g)
    &\defeq
       \max_{i\in[n]}\bigabs{f(i)-g(i)}.
\end{align*}
Since $d_\infty(\,\cdot\,,\,\cdot\,)$ is the only metric we will be using, we
shall simply denote it by $d(\,\cdot\,,\,\cdot\,)$. Observe that for any
$f,g\in S_n$, we have $0 \leq d(f,g) \leq n-1$.

We define the \emph{ball} of radius $r$ centered at $f\in S_n$ as the
set
\begin{align*}
  B_{r,n}(f)
    &\defeq \mathset{ g\in S_n ~|~ d(f,g)\leq r }.
\end{align*}
The infinity metric over $S_n$ is right invariant \cite{DezHua98}, \ie, for
all $f,g,h\in S_n$ we have $d(fh,gh)=d(f,g)$. Thus, the size of a ball depends
only on $r$ and $n$, and not on the choice of the center. We will therefore
denote by $\abs{B_{r,n}}$ the size of a ball of radius $r$ in $S_n$.

For an $n \times n$ matrix, $M=(m_{i,j})$, the \emph{permanent} of $M$ is
defined as
\begin{align*}
  \per(M)
    &\defeq
       \sum_{f\in S_n} \hskip0.5mm
         \prod_{i\in[n]} 
           m_{i,f(i)}.
\end{align*}

\begin{definition}
  A matrix of particular interest is the Toeplitz matrix $A_{r,n}=(a_{i,j})$
  of size $n \times n$ defined by
  \begin{align}
    \label{eq:defa}
    a_{i,j}
      &\defeq
         \begin{cases}
           1 & \abs{i-j} \leq r \\
           0 & \text{otherwise}
         \end{cases},
           \quad i,j \in [n].
  \end{align}
  \defend
\end{definition}

The following lemma is well known~\cite{Klo08,Klo11,TamSch10,Sch09}.

\begin{lemma}
  \label{lemma:relationship:permanent:ball:size:1}
  With the above definitions, it holds that
  \begin{align*}
    \abs{B_{r,n}}
      &= \per(A_{r,n}).
  \end{align*}
\end{lemma}

\begin{IEEEproof}
  This result follows from
  \begin{align*}
    \per(A_{r,n}) 
      &= \sum_{f\in S_n} \prod_{i\in[n]} a_{i,f(i)} \\
      &= \bigabs{\mathset{ f\in S_n ~|~ \forall i \in [n]: 
                             \abs{i-f(i)}\leq r}} \\
      &= \bigabs{\mathset{ f\in S_n ~|~ d(\id_n,f)\leq r}}\\
      &= \bigabs{B_{r,n}(\id_n)}= \abs{B_{r,n}}.
  \end{align*}
\end{IEEEproof}

Note that for any \emph{fixed} radius $r$, tight asymptotic bounds on
$\abs{B_{r,n}}$ are known~\cite{Leh70,Sch09,Klo09,Sta11}. However, in
this paper we are interested in the case of radius $r=\rho \cdot
(n\!-\!1)$, where $\rho\in[0,1]$ is a real constant. This is motivated
by the scaling of rank-modulation schemes. Consider flash-memory cells
as an example, with fixed minimal and maximal charge levels. When
increasing the number of cells, $n$, and assuming a bounded charge
level distortion, we obtain a distance (due to distortion) that grows
linearly with $n$. Note that in expressions like $r = \rho \cdot
(n\!-\!1)$ we always implicitly assume that $\rho$ is such that $r$ is
an integer, and we shall therefore assume throughout the paper that
$\rho$ is in fact rational. We call $\rho$ the \emph{normalized
  radius}.

Because of this particular asymptotic setup, $A_{\rho,n}$ and $B_{\rho,n}$
will in the following, with a slight abuse of notation, stand for $A_{\rho
  \cdot (n-1), n}$ and $B_{\rho \cdot (n-1), n}$, respectively. Moreover,
because $\abs{B_{\rho,n}} = 1$ for $\rho = 0$ and $\abs{B_{\rho,n}} = n!$ for
$\rho = 1$, \ie, the size of balls for $\rho = 0$ and $\rho = 1$ are known
exactly, in the following we will typically focus on the range $0 < \rho < 1$
instead of the range $0 \leq \rho \leq 1$.

We conclude this section by recalling a variety of definitions and results
that we will use throughout the paper.
\begin{itemize}

\item Stirling's approximation of $n!$ (see, \eg,~\cite{GraKnuPat94}) is
  \begin{align*}
    n!
      &= \parenv{\frac{n}{e}}^n \cdot 2^{o(n)}.
  \end{align*}

\item The binary entropy function is defined to be
  \begin{align*}
    h(x) 
      &\defeq 
         - 
         x \cdot \log_2(x) 
         - 
         (1 \! - \! x) \cdot \log_2(1 \! - \!x).
  \end{align*}

\item The Lambert $W$ function is defined by
  \begin{align*}
    z 
      &= W(z) 
           \cdot
           \exp\bigl( W(z) \bigr).
  \end{align*}
  (In this paper, $z$ is limited to non-negative real values.)

\item A \emph{doubly-stochastic matrix} is a square $n\times n$ matrix with
  non-negative real entries for which the sum of each row and each column
  equals $1$.

\item The expression $0 \cdot \log_2(0)$ and the expression $0 \cdot \log_2 \!
  \left( \frac{0}{0} \right)$ are both defined to be equal to $0$.

\end{itemize}

\section{Results}
\label{sec:results:1}

The main results of this paper are new lower bounds on the size of balls over
permutations with the infinity metric. The quality of these new lower bounds
is measured by the asymptotic gap between the known upper bounds and the new
lower bounds. In this section, we first define what we mean by an asymptotic
gap. We then state this gap for known upper and known lower bounds in
Section~\ref{sec:results:known:1} and for known upper and new lower bounds in
Section~\ref{sec:results:new:1}. All derivations for these results will be
given in Sections~\ref{sec:prelim} and~\ref{sec:newbound}.

\begin{definition}
  \label{def:gap:1}
  Fix some real constant $\rho \in (0,1)$. Given some upper bound $\bu$ and
  some lower bound $\bl$ on the ball size, \ie,
  \begin{align*}
    \bl(\rho,n)
      &\leq
         \abs{B_{\rho,n}}
       \leq
         \bu(\rho,n),
  \end{align*}
  where the inequalities are assumed to hold for all positive integers $n$ for
  which $\rho \cdot (n \! - \! 1)$ is an integer, we define the asymptotic gap
  between the upper bound $\bu$ and the lower bound $\bl$ to be
  \begin{align*}
    \gap_{\bl}^{\bu}(\rho)
      &\defeq
        \limsup_{n\rightarrow\infty}
          \frac{1}{n}
            \log_2 \!
              \left(
                \frac{\bu\big( \rho, n \big)}
                     {\bl\big( \rho, n \big)}
              \right).
  \end{align*}
  \defend
\end{definition}

In Section~\ref{sec:implications:1}, we will discuss some ball-packing
and some Gilbert--Varshamov type bounds (see, \eg, \cite[Section
  17.7]{MacSlo78}), both of which depend on the asymptotic size of
balls. Clearly, the better we know the asymptotic size of balls, \ie,
the smaller the gap $\gap_{\bl}^{\bu}(\rho)$ is, the stronger the
statements will be.

\subsection{Gap Based on Known Upper and Known Lower Bounds}
\label{sec:results:known:1}

Based on a known upper bound $\bu_1$ and a known lower bound $\bl_1$,
both detailed in Theorem \ref{th:known:bounds:1}, we obtain the
following result.

\begin{theorem}
  \label{th:results:oldgap:1:1}
  It holds that
  \begin{align*}
    \gap^{\bu_1}_{\bl_1}(\rho)
      &= \begin{cases}
           \big(4 \! - \! 2\log_2(e) \big) \cdot \rho
             &\! 0 < \rho \leq \frac{1}{2} \\[0.50cm]
           2 \cdot (\rho \! - \! 1) \cdot \log_2(e)\\
           \quad
           - 
           (2\rho\! + \! 1) \cdot \log_2(\rho)
             &\! \frac{1}{2} \leq \rho < 1
         \end{cases}
  \end{align*}
  \theoremend
\end{theorem}

The lower bound of $\bl_1$ was very recently improved, and an
asymptotic analysis of this improvement, which we denote $\bl'_1$, is
given in Theorem \ref{th:known:bounds:1:mod}. This improvement,
however, holds only for half the range of $\rho$. By comparing the
upper bound of $\bu_1$ and the improved known lower bound $\bl'_1$, we
obtain the following result.

\begin{theorem}
  \label{th:results:oldgap:1:1:mod}
  For all $0 < \rho < \frac{1}{2}$ it holds that
  \begin{align*}
  \gap^{\bu_1}_{\bl'_1}(\rho)
     &= 2
           \cdot 
           \big( 
             h(\mu^*) 
             + 
             \log_2(\mu^*) 
           \big)
           \cdot
           \rho,
  \end{align*}
  where $\mu^*$ is the constant defined by
  \begin{align*}
    \mu^* 
      &\defeq 
         \frac{1}{1+W(e^{-1})}
       \approx 
         0.782.
  \end{align*}
  \theoremend
\end{theorem}

The result of Theorems~\ref{th:results:oldgap:1:1}
and~\ref{th:results:oldgap:1:1:mod} are visualized by curves (a) and (a'),
respectively, in Fig.~\ref{fig:bounds}.

\subsection{Gap Based on Known Upper and New Lower Bounds}
\label{sec:results:new:1}

We present two new lower bounds on the size of balls over permutations with
the infinity metric. The first new lower bound, denoted $\bl_2$, is given in
Theorem \ref{th:newlow:2}. The other new lower bound, denoted $\bl_3$, is
detailed in Theorems~\ref{th:newlow:3:1}
and~\ref{th:lambert:function:based:bound}. Based on a known upper bound
$\bu_1$ and the new lower bounds $\bl_2$ and $\bl_3$, we obtain the following
results.

\begin{theorem}
  \label{th:results:newgap:1:2}
  It holds that
  \begin{align*}
    \gap^{\bu_1}_{\bl_2}(\rho)
      &= \begin{cases}
           \big( 3 \! - \! 2\log_2(e) \big) \cdot \rho
             & 0 < \rho \leq \frac{1}{2} \\[0.50cm]
           2 \cdot (1 \! - \! \rho) \cdot (1 \! - \! \rho-\log_2 e) \\
             \quad -2\rho \cdot \log_2\rho
             & \frac{1}{2} \leq \rho < 1
         \end{cases}
  \end{align*}
  \theoremend
\end{theorem}

\begin{theorem}
  \label{th:results:newgap:1:3}
  It holds that
  \begin{align*}
    \gap^{\bu_1}_{\bl_3}(\rho)
      &= \begin{cases}
           \log_2
             \left(
               \frac{4}{e \cdot \log_2(e)}
             \right)
             & 0 < \rho \leq \frac{1}{2} \\[0.50cm]
           \log_2
             \left(
               \frac{\that}
                    {\log_2(e)}
             \right)
           -
           \that(2\rho-1) \\
           \quad
           -
           \log_2(1-\rho) \\
           \quad
           -
           2(1-\rho) \cdot \log_2(e) \\
           \quad
           -
           2 \rho \cdot \log_2(\rho)
             & \frac{1}{2} < \rho < 1
         \end{cases}
  \end{align*}
  where
  \begin{align*}
    \that
      &\defeq
         \log_2(e)
         \cdot
         \parenv{\frac{2 (1 \! - \! \rho)}
                      {2\rho-1}
         -
         W \!
           \parenv{ \! \frac{(1 \! - \! \rho)
                          \cdot \exp\parenv{\frac{2 (1-\rho)}{2\rho-1}}}
                       {2\rho-1}} \! }.
  \end{align*}
  \theoremend
\end{theorem}

\begin{figure}
  \psfrag{xax} {(a)}
  \psfrag{xatx}{(a')}
  \psfrag{xbx} {(b)}
  \psfrag{xcx} {(c)}
  \psfrag{rho} {$\rho$}
  \psfrag{gap} {$\gap$}
  \centering
  \includegraphics[scale=0.75]{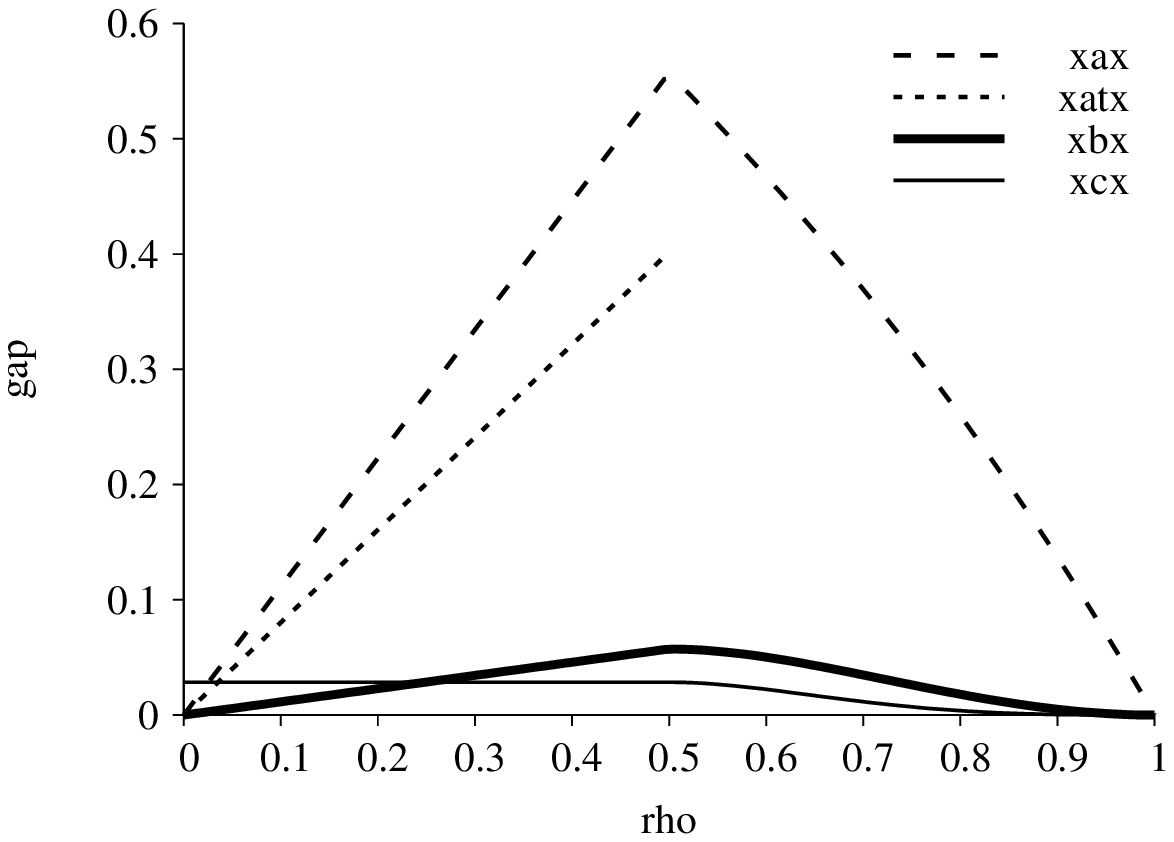}
  \caption{
  (a) $\gap^{\bu_1}_{\bl_1}(\rho)$ of Theorem~\ref{th:results:oldgap:1:1};
  (a') $\gap^{\bu_1}_{\bl'_1}(\rho)$ of Theorem~\ref{th:results:oldgap:1:1:mod};
  (b) $\gap^{\bu_1}_{\bl_2}(\rho)$ of Theorem~\ref{th:results:newgap:1:2};
  (c) $\gap^{\bu_1}_{\bl_3}(\rho)$ of
  Theorem~\ref{th:results:newgap:1:3}.}
  \label{fig:bounds}
\end{figure}

The result of Theorems~\ref{th:results:newgap:1:2}
and~\ref{th:results:newgap:1:3} are visualized by curves (b) and (c),
respectively, in Fig.~\ref{fig:bounds}. Note that the curves (b) and (c) cross
at
\begin{align*}
  \rho 
    &= \frac{\log_2
               \left(
                 \frac{4}{e \cdot \log_2(e)}
               \right)}
            {3 \! - \! 2\log_2(e)}
     \approx 
       0.249.
\end{align*}

\section{Analysis of Known Bounds}
\label{sec:prelim}

The following theorems summarize, to the best of our knowledge, the tightest
known bounds for balls in $(S_n, d_\infty)$.

\begin{theorem}
  \label{th:known:bounds:1}
  It holds that
  \begin{align*}
    \bl_1(\rho,n)
      &\leq 
         \abs{B_{\rho,n}} 
       \leq 
         \bu_1(\rho,n),
  \end{align*}
  where
  \begin{align*}
    &
    \log_2 \bl_1(\rho,n) \\
      &\defeq
         \begin{cases}
           n \cdot \log_2(n) \\
           \quad
           -
           n
           \cdot
           \big[
             \log_2(e)
             -
             1
             + 
             2 \rho
             -
             \log_2(\rho) 
           \big] \\
           \quad
           + 
           o(n)
             & 0 < \rho \leq \frac{1}{2} \\[0.50cm]
           n \cdot \log_2(n) \\
           \quad
           -
           n
           \cdot
           \big[
             \log_2(e)
             -
             \log_2(\rho)
           \big] \\
           \quad
           + 
           o(n)
             & \frac{1}{2} \leq \rho < 1
         \end{cases}
  \end{align*}
  and
  \begin{align*}
    &
    \log_2 \bu_1(\rho,n) \\
      &\defeq
         \begin{cases}
           n \cdot \log_2(n) \\
           \quad
           -
           n
           \cdot
           \big[
             \big( \log_2(e) \! - \! 1 \big) \cdot (2\rho \! + \! 1)
             -
             \log_2(\rho)
           \big] \\
           \quad
           +
           o(n)
           & 0 < \rho \leq \frac{1}{2} \\[0.50cm]
           n \cdot \log_2(n) \\
           \quad
           -
           n
           \cdot
           \big[
             \log_2(e) \cdot (3-2\rho)
             +
             2\rho \cdot \log_2(\rho)
           \big] \\
           \quad
           +
           o(n)
             & \frac{1}{2} \leq \rho < 1
         \end{cases}
  \end{align*}
\end{theorem}

\begin{IEEEproof}
  These bounds follow from results
  in~\cite{Klo11,Klo08,TamSch10,FarSchBru16a}. For more details, see
  Appendix~\ref{sec:proof:th:known:bounds:1}.
\end{IEEEproof}

\medskip

Comparing the upper bound $\bu_1$ with the lower bound $\bl_1$, one obtains
immediately the result stated in Theorem~\ref{th:results:oldgap:1:1}.

\begin{theorem}
  \label{th:known:bounds:1:mod}
  For all $0 < \rho \leq \frac{1}{2}$, it holds that
  \begin{align*}
    \abs{B_{\rho,n}}
      &\geq
         \bl'_1(\rho,n),
  \end{align*}
  where
  \begin{align*}
    \log_2 \bl'_1(\rho,n)
      &= n \cdot \log_2 n \\
      &\quad\quad
         - 
         n 
           \cdot
           \big[
             \big( \log_2(e) \! - \! 1 \big) \cdot (2\rho \! + \! 1)
             -
             \log_2(\rho) \\
      &\quad\quad\quad\quad\quad\quad
             +
             2
               \cdot 
               \big(h(\mu^*) \! + \! \log_2(\mu^*) \big)
               \cdot
               \rho
           \big] \\
      &\quad\quad
           +
           o(n),
  \end{align*}
  and where $\mu^*$ is the constant defined by
  \begin{align*}
     \mu^* 
       &\defeq 
          \frac{1}{1+W(e^{-1})}
        \approx 
          0.782.
  \end{align*}
\end{theorem}

\begin{IEEEproof}
  This result follows from an asymptotic analysis of the conjectured lower
  bound in~\cite{Klo11}, which was very recently proven in
  \cite{GuoYan16}. The asymptotic analysis is briefly sketched in
  Appendix~\ref{sec:proof:th:known:bounds:1:mod}.
\end{IEEEproof}

\section{New Lower Bounds}
\label{sec:newbound}

In this section we present new lower bounds on the size of balls in $(S_n,
d_\infty)$. These lower bounds are based on the following theorem, which is a
variant of a result in~\cite{LinSamWig00}. The art in using this theorem is to
find $Q$ matrices that yield large right-hand sides
in~\eqref{eq:th:sinkhornperlower} and that are analytically tractable.

\begin{theorem}
  \label{th:sinkhornperlower}
  Let $M \defeq (m_{i,j})$ be an $n \times n$ matrix with non-negative
  entries and $\per(M) > 0$, and let $Q \defeq (q_{i,j})$ be an $n
  \times n$ doubly-stochastic matrix such that $q_{i,j}=0$ whenever
  $m_{i,j}=0$. Then
  \begin{align}
    \log_2\per(M)
      &\geq
         \log_2 \!
           \left(
             \frac{n!}{n^n}
           \right)
         +
         \sum_{i,j \in [n]}
           \parenv{-q_{i,j}\log_2 \frac{q_{i,j}}{m_{i,j}}}.
             \label{eq:th:sinkhornperlower}
  \end{align}
\end{theorem}

\begin{IEEEproof}
  Let $\varepsilon > 0$ and let $M^{(\varepsilon)} \defeq \big(
  m^{(\varepsilon)}_{i,j} \big)$ be the matrix that is obtained from $M$ by
  replacing zeros by $\varepsilon$. Because $M^{(\varepsilon)}$ contains only
  strictly positive entries, it follows from a theorem by
  Sinkhorn~\cite{Sin64} that there exist two diagonal matrices $D$ and
  $D'$ with positive diagonal elements such that $D \cdot M^{(\varepsilon)}
  \cdot D'$ is a doubly-stochastic matrix.  Let $D$ and $D'$ be given by
  \begin{align*}
    D
      &\defeq
         \diag(d_1, \ldots, d_n), \\
    D'
      &\defeq
         \diag(d'_1, \ldots, d'_n),
  \end{align*}
  where $d_i$, $i \in [n]$, and $d'_j$, $j \in [n]$, are positive real
  numbers. Note that the element in the $i$-th row and the $j$-th column of $D
  \cdot M^{(\varepsilon)} \cdot D'$ is given by $d_i \cdot
  m^{(\varepsilon)}_{i,j} \cdot d'_j$. Then
  \begin{align}
    &\!\!
    \log_2
      \per\big( M^{(\varepsilon)} \big) \nonumber \\
      &\overset{\mathrm{(a)}}{=} 
         \log_2 \per\big( D \cdot M^{(\varepsilon)} \cdot D' \big)
         -
         \sum_{i \in [n]}
           \log_2(d_i)
         -
         \sum_{j \in [n]}
           \log_2(d'_j) \nonumber \\
       &\overset{\mathrm{(b)}}{\geq}
          \log_2 \!
           \left(
             \frac{n!}{n^n}
           \right)
           -
         \sum_{i \in [n]}
           \log_2(d_i)
         -
         \sum_{j \in [n]}
           \log_2(d'_j) \nonumber \\
       &\overset{\mathrm{(c)}}{\geq}
          \log_2 \!
            \left(
              \frac{n!}{n^n}
            \right)
          -
         \sum_{i \in [n]}
           \log_2(d_i)
         -
         \sum_{j \in [n]}
           \log_2(d'_j) \nonumber \\
       &\quad\quad
          -
          \sum_{i,j \in [n]}
            q_{i,j}
            \cdot
            \log_2 \!
              \left(
                \frac{q_{i,j}}
                     {d_i \cdot m^{(\varepsilon)}_{i,j} \cdot d'_j}
              \right) \nonumber \\
       &= \log_2 \!
            \left(
              \frac{n!}{n^n}
            \right)
          -
          \sum_{i,j \in [n]}
            q_{i,j}
            \cdot
            \log_2 \!
              \left(
                \frac{q_{i,j}}
                     {m^{(\varepsilon)}_{i,j}}
              \right) \nonumber \\
       &\overset{\mathrm{(d)}}{\geq}
         \log_2 \!
            \left(
              \frac{n!}{n^n}
            \right)
          -
          \sum_{i,j \in [n]}
            q_{i,j}
            \cdot
            \log_2 \!
              \left(
                \frac{q_{i,j}}
                     {m_{i,j}}
              \right).
                \label{eq:th:sinkhornperlower:proof:1}
  \end{align}
  We justify the steps taken: Step~$\mathrm{(a)}$ follows by noting
  that $M^{(\varepsilon)}$ is obtained from $D\cdot M^{(\varepsilon)}
  \cdot D'$ by factoring out $d_i$ from the $i$th row and $d'_j$ from
  the $j$th column, for all $i,j\in[n]$. Step~$\mathrm{(b)}$ follows
  from Van der Waerden's conjecture (proven by Falikman~\cite{Fal81}
  and by Egorychev~\cite{Ego81}), which states that for any $n\times
  n$ doubly-stochastic matrix $U$, we have $\per(U)\geq
  n!/n^n$. Step~$\mathrm{(c)}$ follows from the non-negativity of
  relative entropy. Finally, Step~$\mathrm{(d)}$ follows from
  $m^{(\varepsilon)}_{i,j} \geq m_{i,j}$ for all $i,j \in [n]$, and by
  noting that we require $q_{i,j}=0$ whenever $m_{i,j}=0$.
  
  Note that $\lim_{\varepsilon \to 0} \per\bigl( M^{(\varepsilon)} \bigr) =
  \per(M)$, because the permanent of a matrix is a multilinear function of the
  entries of the matrix. With this, applying the limit $\varepsilon \to 0$ to
  the expression in~\eqref{eq:th:sinkhornperlower:proof:1}, we
  obtain~\eqref{eq:th:sinkhornperlower}.
\end{IEEEproof}

We note that $D$ and $D'$ are auxiliary matrices in the proof of
Theorem~\ref{th:sinkhornperlower}. Only their existence matters, while their
entries do not play a role in \eqref{eq:th:sinkhornperlower}.
For matrices $M$ with strictly positive entries (and possibly some
other classes of matrices), the right-hand side
of~\eqref{eq:th:sinkhornperlower} can be maximized with the help of
Sinkhorn's balancing algorithm~\cite{Sin64}, see, \eg, the discussions
in~\cite{LinSamWig00, HuaJeb09, Von14}.\footnote{Strictly speaking, we
  do not need Sinkhorn's balancing algorithm to prove our
  results. However, we mention this algorithm here because it played a
  key role for analyzing setups with finite $n$ and $r$ and coming up
  with the $Q_{r,n}$ matrices in
  Definitions~\ref{def:newlow:3:1:matr:Q}
  and~\ref{def:newlow:3:2:matr:Q} for general $n$ and $r$.}

In the following, we will apply Theorem~\ref{th:sinkhornperlower} with $M =
A_{r,n}$ and with two classes of $Q$ matrices. The first class of $Q$ matrices
will ultimately yield Theorem~\ref{th:results:newgap:1:2}, whereas the second
class of $Q$ matrices will ultimately yield
Theorem~\ref{th:results:newgap:1:3}.

\begin{figure}
  \centering
  \subfigure[$Q_{6,20}$ of Def.~\ref{def:first:class:Q:matrices}]
            {\includegraphics[scale=0.20]{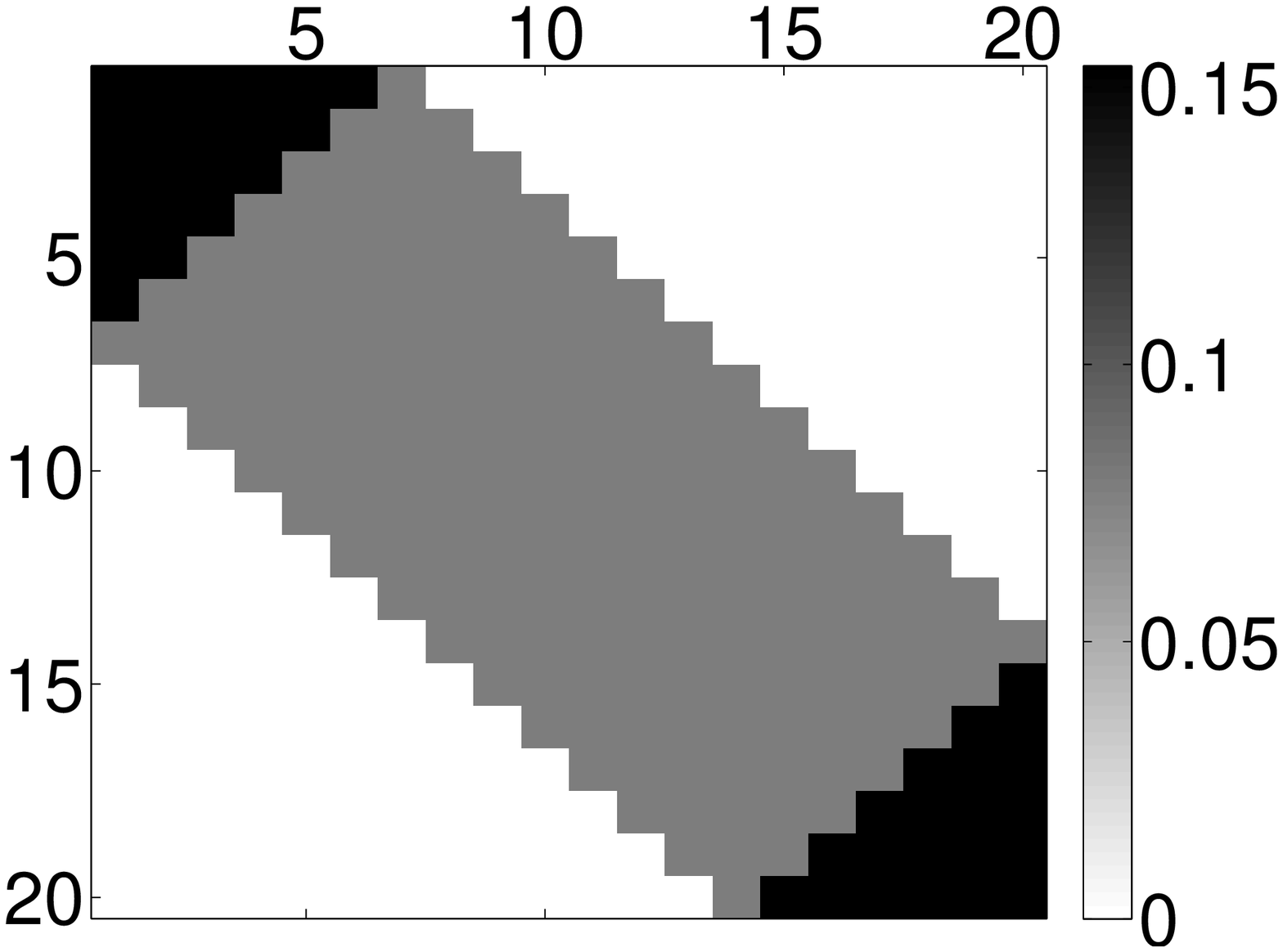}}
  \subfigure[$Q_{14,20}$ of Def.~\ref{def:first:class:Q:matrices}]
            {\includegraphics[scale=0.20]{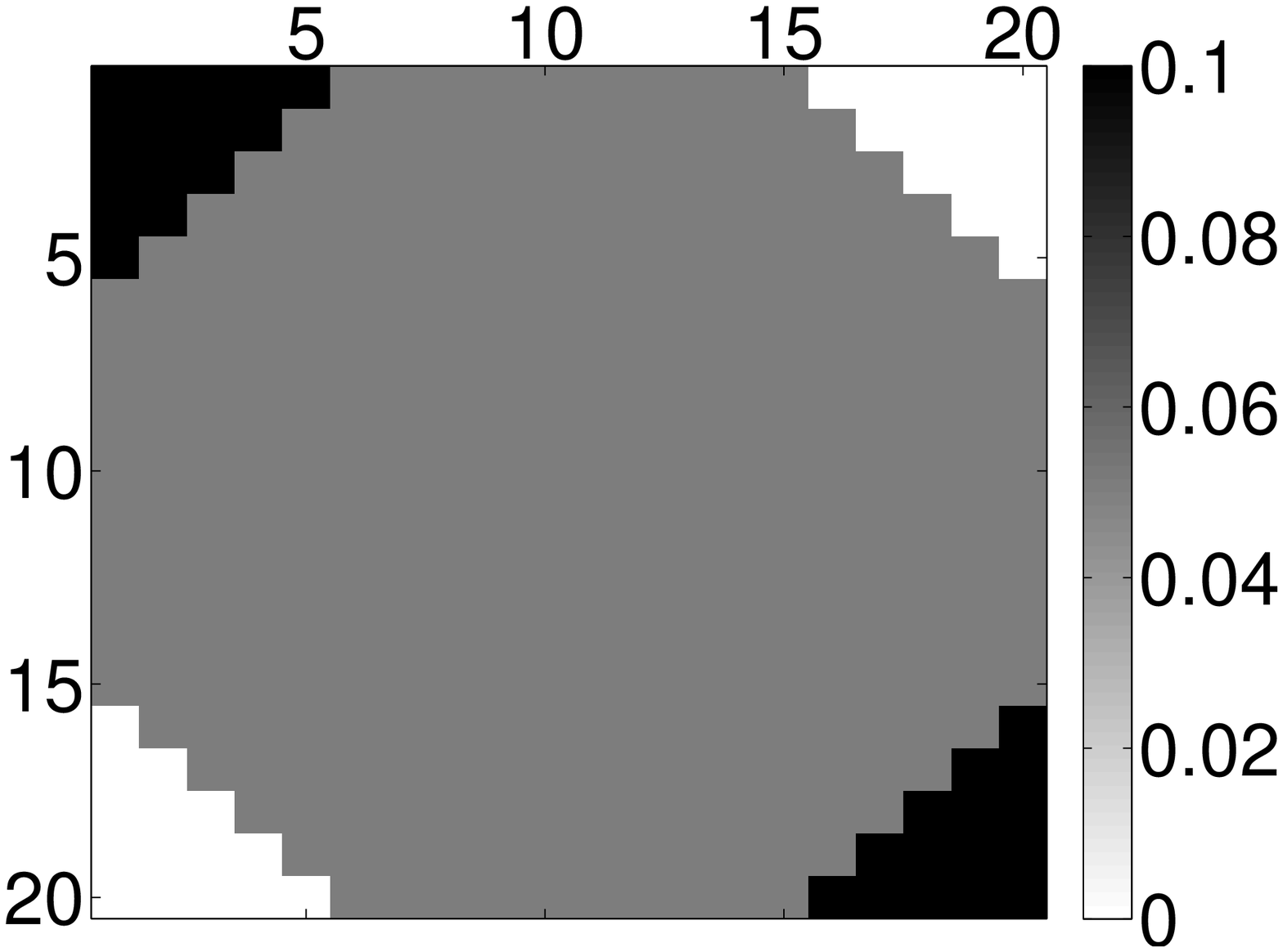}}

  \subfigure[$Q_{6,20}$ of Def.~\ref{def:newlow:3:1:matr:Q}]
            {\includegraphics[scale=0.20]{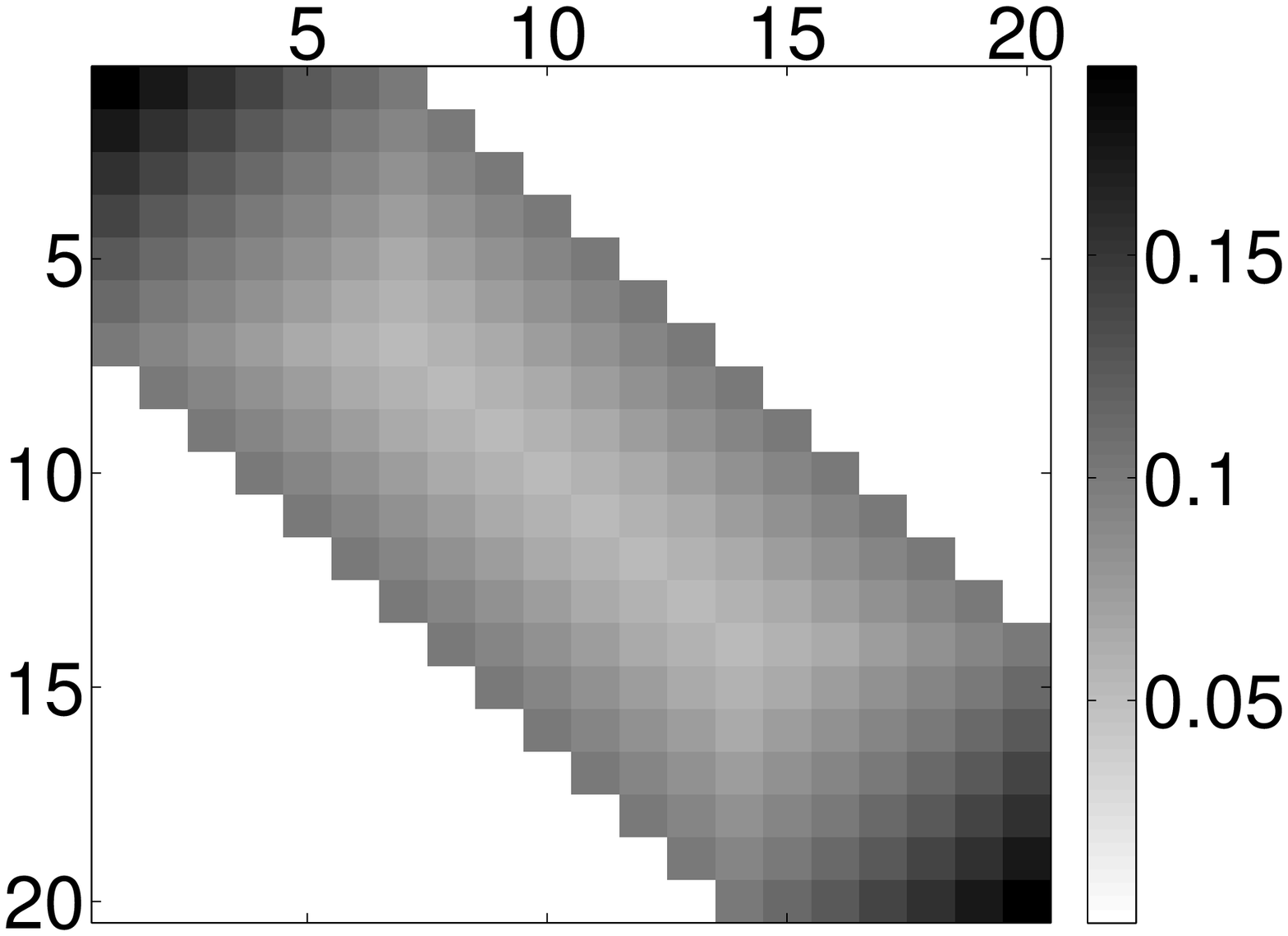}}
  \subfigure[$Q_{14,20}$ of Def.~\ref{def:newlow:3:2:matr:Q}]
            {\includegraphics[scale=0.20]{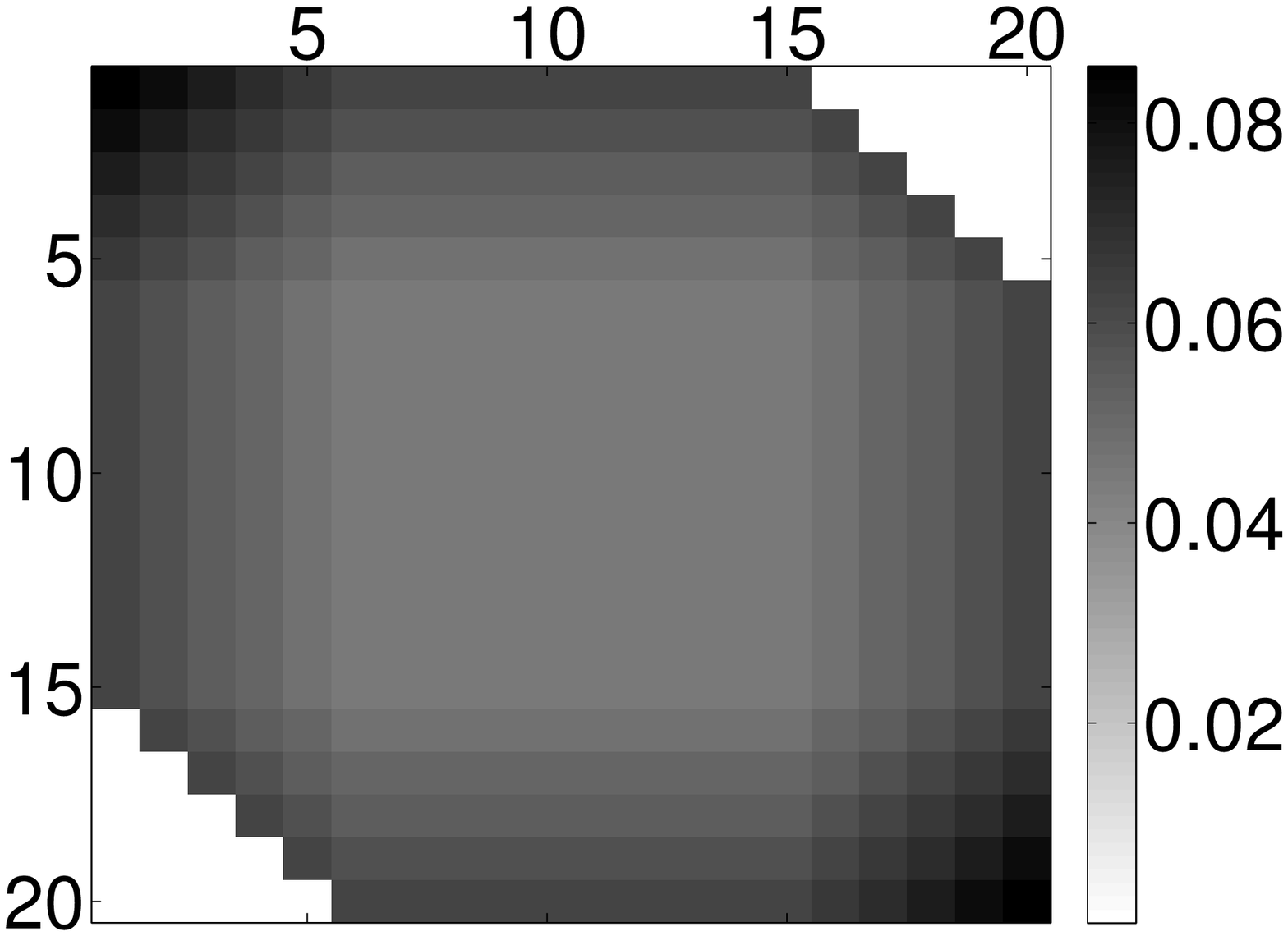}}
  \caption{Matrices $Q_{r,n}$ used in this paper to obtain various
    lower bounds. Note that for every instance, the support of
    $Q_{r,n}$ equals the support of $A_{r,n}$.}
  \label{fig:q:matrix}
\end{figure}

\subsection{First Class of $Q$ Matrices}
\label{sec:first:class:Q:matrices:1}

\begin{definition}
  \label{def:first:class:Q:matrices}
  For $0 \leq r \leq \frac{n-1}{2}$, we define the matrix $Q_{r,n} =
  (q_{i,j})$ with entries
  \begin{equation}
    \label{eq:first}
    q_{i,j}
      \defeq
         \begin{cases}
           \frac{2}{2r+1}      
             & \text{$i+j\leq r+1$ \ or \ $i+j\geq 2n-r+1$} \\[0.25cm]
           \frac{a_{i,j}}{2r+1} 
             & \text{otherwise}
         \end{cases}.
  \end{equation}
  For $\frac{n-1}{2} \leq r \leq n-1$, we define the matrix $Q_{r,n} = (q_{i,j})$
  with entries
  \begin{equation}
    \label{eq:second}
    q_{i,j}
      \defeq
         \begin{cases}
           \frac{2}{n}      
             & \text{$i+j \leq n-r$ \ or \ $i+j \geq n+r+2$} \\[0.25cm]
           \frac{a_{i,j}}{n} 
             & \text{otherwise}.
         \end{cases}.
  \end{equation}
  \defend
\end{definition}

\begin{lemma}
  \label{lemma:firstQ}
  The matrix $Q_{r,n}$ in Definition \ref{def:first:class:Q:matrices}
  is a doubly-stochastic matrix with the same support as $A_{r,n}$.
\end{lemma}
\begin{IEEEproof}
  See
  Appendix~\ref{sec:proof:lemma:firstQ}.
\end{IEEEproof}

For $n = 20$ and $r = 6$, the resulting $Q_{r,n}$ matrix is depicted in
Fig.~\ref{fig:q:matrix}(a), whereas for $n = 20$ and $r = 14$, the resulting
$Q_{r,n}$ matrix is depicted in Fig.~\ref{fig:q:matrix}(b).

\begin{theorem}
  \label{th:newlow:2}
  Fix some $\rho$, $0 < \rho < 1$. It holds that
  \begin{align*}
    \abs{B_{\rho,n}}
      &\geq \bl_2(\rho,n),
  \end{align*}
  where we define
  \begin{align*}
    &
    \log_2 \bl_2(\rho,n) \\
      &\defeq
         \begin{cases}
           n \cdot \log_2(n) \\
           \quad
           -
           n
             \cdot
             \big[
               \log_2(e)
               -
               1
               +
               \rho
               -
               \log_2(\rho)
             \big] \\
           \quad
           +
           o(n)
             & 0 < \rho \leq \frac{1}{2} \\[0.5cm]
          n \cdot \log_2(n) \\
          \quad
          -
          n
            \cdot
            \big[
              \log_2(e)
              +
              2
              \cdot
              (1-\rho)^2
            \big] \\
          \quad
          +
          o(n)
            & \frac{1}{2} \leq \rho < 1
        \end{cases}
  \end{align*}
\end{theorem}

\begin{IEEEproof}
  See Appendix~\ref{sec:proof:th:newlow:2}.
\end{IEEEproof}

Comparing the upper bound $\bu_1$ with the new lower bound $\bl_2$, one
obtains immediately the result stated in Theorem~\ref{th:results:newgap:1:2}.

We conclude this section with several comments on the matrices
$Q_{r,n}$.  The matrix $Q_{r,n}$ that we defined in \eqref{eq:first}
already appeared in~\cite{Klo11,GuoYan16}, and that of
\eqref{eq:second} already appeared
in~\cite{FarSchBru16a}. Although~\cite{Klo11,GuoYan16,FarSchBru16a}
introduce the same matrices, they consider a different approach to
obtain a lower bound on $\per(A_{r,n})$ than the one presented in this
paper, and with that they obtain a different lower bound on
$\abs{B_{\rho,n}}$.

We also note that the definition of the matrix $Q_{r,n}$ in
\eqref{eq:first} works not only for the range $0 < \rho \leq
\frac{1}{2}$, but for the entire range $0 < \rho < 1$. However, we
define these matrices only for the range $0 < \rho \leq \frac{1}{2}$,
because they yield a weaker lower bound than the lower bound in
Theorem~\ref{th:newlow:2} for the range $\frac{1}{2} \leq \rho < 1$.

\subsection{Second Class of $Q$ Matrices}
\label{sec:second:class:Q:matrices:1}

Our second class for the $Q$ matrices are more sophisticated than our first
class. In the following, we will separately discuss the cases $0 < \rho \leq
1/2$ and $\frac{1}{2} < \rho < 1$. Note that the $Q$ matrix that we will
use for the case $\frac{1}{2} < \rho < 1$ maximizes the right-hand side
of~\eqref{eq:th:sinkhornperlower} for $M = A_{r,n}$. This is in contrast to
the $Q$ matrix that we will use for the case $0 < \rho \leq 1/2$. This $Q$
matrix does not, in general, maximize the right-hand side
of~\eqref{eq:th:sinkhornperlower} for $M = A_{r,n}$.  An exception is the case
where $n$ is even and $r = \frac{n-2}{2}$.

\subsubsection{Range $0 < \rho \leq \frac{1}{2}$}

We start our discussion of this case with the following definition.

\begin{definition}
  \label{def:newlow:3:1:matr:Q}
  Fix some $r$, $1 \leq r \leq \frac{n-2}{2}$. Let $Q_{r,n} \defeq
  (q_{i,j})$ be the $n \times n$-matrix with entries
  \begin{align*}
    q_{i,j} 
      &\defeq 
         a_{i,j} \cdot C \cdot \tilde q_{i,j} \ ,
  \end{align*}
  where
  \begin{align*}
    \tilde q_{i,j}
      = \begin{cases}
          \alpha_r^{(r+1-i)+(r+1-j)}
            & \text{$1 \leq i \leq r+1$, \
                    $1 \leq j \leq r+1$} \\[0.25cm]
          \alpha_r^{i-(n-r)+j-(n-r)}
            & \text{$n \! - \! r \leq i \leq n$, \
                    $n \! - \! r \leq j \leq n$} \\[0.25cm]
          \alpha_r^{|i-j|}
            & \text{otherwise}
        \end{cases},
  \end{align*}  
  where $C$ is given by
  \begin{align}
    C
      &\defeq 
         (\alpha_r - 1) \cdot \alpha_r^{-r-1}
       = \frac{\alpha_r - 1}{\alpha_r + 1},
           \label{eq:def:newlow:3:1:C}
  \end{align}
  and where $\alpha_r > 0$ satisfies
  \begin{align}
    \alpha_r^{r+1} - \alpha_r - 1 
      &= 0.
           \label{eq:def:newlow:3:1:alpha}
  \end{align}
  \defend
\end{definition}

Note that the second expression for $C$ in~\eqref{eq:def:newlow:3:1:C} follows
from the first expression for $C$ in~\eqref{eq:def:newlow:3:1:C}, along
with~\eqref{eq:def:newlow:3:1:alpha}. Moreover, note that the $\alpha_r > 0$
satisfying~\eqref{eq:def:newlow:3:1:alpha} is unique. (This can be proven by
analyzing the function $\alpha \mapsto \alpha^{r+1} - \alpha - 1$.)

For $n = 20$ and $r = 6$, the resulting $Q_{r,n}$ matrix is depicted in
Fig.~\ref{fig:q:matrix}(c).

\begin{lemma}
  \label{lemma:newlow:3:1:Q:properties}
  The matrix $Q_{r,n}$ in Definition~\ref{def:newlow:3:1:matr:Q} is a
  doubly-stochastic matrix with the same support as $A_{r,n}$.
\end{lemma}

\begin{IEEEproof}
  See
  Appendix~\ref{sec:proof:lemma:newlow:3:1:Q:properties}.
\end{IEEEproof}

\begin{theorem}
  \label{th:newlow:3:1}
  Fix some $\rho$, $0 < \rho \leq \frac{1}{2}$. It holds that
  \begin{align*}
    \abs{B_{\rho,n}}
      &\geq \bl_3(\rho,n),
  \end{align*}
  where we define
  \begin{align*}
    &
    \log_2
      \bl_3(\rho,n) \\
      &\defeq
         n \cdot \log_2(n) \\
      &\quad\ 
         -
         n
         \cdot 
           \Big[
             \big( \log_2(e) \! - \! 1 \big) \cdot 2\rho
             - 
             \log_2(\rho)
             -
             \log_2\bigl( \log_2(e) \bigr)
             +
             1
           \Big] \\
       &\quad\  
         +
         o(n).
  \end{align*}
\end{theorem}

\begin{IEEEproof}
  See Appendix~\ref{sec:proof:th:newlow:3:1}.
\end{IEEEproof}

Comparing the upper bound $\bu_1$ with the new lower bound $\bl_3$, one
obtains immediately the result stated in Theorem~\ref{th:results:newgap:1:3}
for the case $0 < \rho \leq \frac{1}{2}$.

\subsubsection{Range $\frac{1}{2} < \rho < 1$}

We start our discussion of this case with the following definition.

\begin{definition}
  \label{def:newlow:3:2:matr:Q}
  Fix some $r$, $\frac{n-1}{2} < r < n \! - \! 1$. Let $Q_{r,n}
  \defeq (q_{i,j})$ be the $n \times n$-matrix with entries
  \begin{align}
    q_{i,j}
      &= a_{i,j} \cdot C \cdot \exp_2(\lambda_i) \cdot \exp_2(\lambda'_j),
           \quad i,j \in [n],
             \label{eq:def:newlow:3:2:matr:Q}
  \end{align}
  where 
  \begin{align}
    \lambda_i
      &\defeq 
         \begin{cases}
           \big( ( n \! - \! r) - i \big) \cdot \log_2(\alpha_{r,n})
               & 1 \leq i \leq n-r \\[0.10cm]
           0
               & n -r \leq i \leq r+1 \\[0.10cm]
           \big( i - (r \! + \! 1) \big) \cdot \log_2(\alpha_{r,n})
               & r+1 \leq i \leq n
         \end{cases}, \nonumber \\
    \lambda'_j
      &\defeq
         \lambda_j\ ,
           \quad j\in[n], \nonumber \\
      C
        &\defeq
           (\alpha_{r,n} \! - \! 1) 
           \cdot
           \alpha_{r,n}^{-(n-r)} 
             \label{eq:def:newlow:3:2:C:1} \\
        &= \frac{\alpha_{r,n} \! - \! 1} 
                {(2r \! - \! n \! + \! 2) 
                 - (2r \! - \! n) \cdot \alpha_{r,n}} \ ,
             \label{eq:def:newlow:3:2:C:2}
  \end{align}
  where $\alpha_{r,n} > 0$ satisfies
  \begin{align}
    \alpha_{r,n}^{n-r}
    +
    (2r \! - \! n)
    \cdot
    \alpha_{r,n}
    -
    (2r \! - \! n \! + \! 2)
      &= 0.
           \label{eq:def:newlow:3:2:alpha}
  \end{align}
  \defend
\end{definition}

We note that
  \begin{itemize}
  
  \item $\lambda_{n+1-i} = \lambda_i$\ , $i\in[n]$, 

  \item $\lambda'_{n+1-j} = \lambda'_j$\ , $j\in[n]$.

  \end{itemize}
Additionally, we observe that the second expression for $C$
in~\eqref{eq:def:newlow:3:2:C:2} follows from the first expression for
$C$ in~\eqref{eq:def:newlow:3:2:C:1}, along
with~\eqref{eq:def:newlow:3:2:alpha}. Moreover, note that the
$\alpha_{r,n} > 0$ satisfying~\eqref{eq:def:newlow:3:2:alpha} is
unique. (This can be proven by analyzing the function $\alpha \mapsto
\alpha^{n-r} + (2r \!  - \!  n) \cdot \alpha - (2r \! - \! n \! + \!
2)$.)

For $n = 20$ and $r = 14$, the resulting $Q_{r,n}$ matrix is depicted in
Fig.~\ref{fig:q:matrix}(d).

\begin{lemma}
  \label{lemma:newlow:3:2:matr:Q}
  The matrix $Q_{r,n}$ in Definition~\ref{def:newlow:3:2:matr:Q} is a
  doubly-stochastic matrix with the same support as $A_{r,n}$.
\end{lemma}

\begin{IEEEproof}
  See Appendix~\ref{sec:proof:lemma:newlow:3:2:matr:Q}.
\end{IEEEproof}

\medskip

\begin{lemma}
  \label{lemma:newlow:3:2:partial:result:1}
  Fix some $r$, $\frac{n-1}{2} < r < n \! - \! 1$. It holds that
  \begin{align*}
    \log_2 \abs{B_{r,n}}
      &\geq
         \log_2(n!)
         -
         n
           \log_2(n)
         -
         n
           \cdot
           \log_2(\alpha_{r,n} \! - \! 1) \\
      &\quad\ 
         +
         (n-r)
         \cdot
           (2r-n+2)
           \cdot
           \log_2(\alpha_{r,n}),
  \end{align*}
  where $\alpha_{r,n}$ was specified in~\eqref{eq:def:newlow:3:2:alpha}.
\end{lemma}

\begin{IEEEproof}
  See Appendix~\ref{sec:proof:lemma:newlow:3:2:partial:result:1}.
\end{IEEEproof}

\medskip

Note that the lower bound in Lemma~\ref{lemma:newlow:3:2:partial:result:1}
contains the constant $\alpha_{r,n}$. In order to get rid of this constant,
the upcoming Lemma~\ref{lemma:newlow:3:2:alpha:simplification:1} suitably
approximates this constant and Theorem~\ref{th:lambert:function:based:bound}
will then show the updated expression for the lower bound.

\begin{lemma}
  \label{lemma:newlow:3:2:alpha:simplification:1}
  Fix some $\rho$ with $\frac{1}{2} <\rho < 1$. Let $r \defeq \rho
  \cdot (n\!-\!1)$. Then $\alpha_{r,n}$ from~\eqref{eq:def:newlow:3:2:alpha}
  satisfies
  \begin{align*}
    \alpha_{r,n}
      &= 1
         +
         \parenv{\that \! + \! \Theta\big( n^{-1} \big)}
           \cdot
           \parenv{2^{\frac{1}{(n\!-\!1)(1-\rho)+1}} \! - \! 1},
  \end{align*}
  where
  \begin{align}
    \label{eq:defthat}
    \that
      &\defeq
         \log_2(e)
         \cdot
         \parenv{\frac{2 (1 \! - \! \rho)}
                      {2\rho-1}
         -
         W \!
           \parenv{ \! \frac{(1 \! - \! \rho)
                          \cdot \exp\parenv{\frac{2 (1-\rho)}{2\rho-1}}}
                       {2\rho-1}} \! }.
  \end{align}
\end{lemma}

\begin{IEEEproof}
  See Appendix~\ref{sec:proof:lemma:newlow:3:2:alpha:simplification:1}.
\end{IEEEproof}

\medskip

\begin{theorem}
  \label{th:lambert:function:based:bound}
  Fix some $\rho$ with $\frac{1}{2} < \rho < 1$ a constant. Let $r \defeq \rho
  \cdot (n\!-\!1)$. It holds that
  \begin{align*}
    \abs{B_{r,n}}
      &\geq
         \bl_3(\rho,n),
  \end{align*}
  where
  \begin{align*}
    &
    \log_2 \bl_3(\rho,n) \\
      &\quad\defeq 
         n \cdot \log_2(n) \\
      &\quad\quad\ 
         -
         n
         \cdot
           \left[
             \log_2 \!
               \left(
                 \frac{e \cdot \that}{\log_2(e)}
               \right)
             -
             \that \cdot (2\rho \! - \! 1)
             -
             \log_2(1 \! - \! \rho)
           \right] \\
       &\quad\quad\ 
         +
         o(n),
  \end{align*}
  and where $\that$ is given by \eqref{eq:defthat}.
\end{theorem}

\begin{IEEEproof}
  See Appendix~\ref{sec:proof:th:lambert:function:based:bound}.
\end{IEEEproof}

\medskip

Comparing the upper bound $\bu_1$ with the new lower bound $\bl_3$, one
obtains immediately the result stated in Theorem~\ref{th:results:newgap:1:3}
for the case $\frac{1}{2} < \rho < 1$.

\section{Conclusion and Outlook}
\label{sec:conc}

We conclude this paper by commenting about the newly obtained lower bounds and
by stating some open problems.

\subsection{Implications of the New Bounds}
\label{sec:implications:1}

\begin{figure}
  \psfrag{xex}{\parbox{1cm}{\hskip0.4cm (a)}}
  \psfrag{xfx}{\parbox{1cm}{\hskip-1.1cm (b)}}
  \psfrag{xgx}{\parbox{1cm}{\vskip0.2cm\hskip0.2cm (c)}}
  \psfrag{delta}{\parbox{2cm}{\hskip0.2cm $\delta$}}
  \psfrag{rrr}{\parbox{1cm}{\vskip0.7cm $R$}}
  \centering
  \includegraphics[scale=0.7]{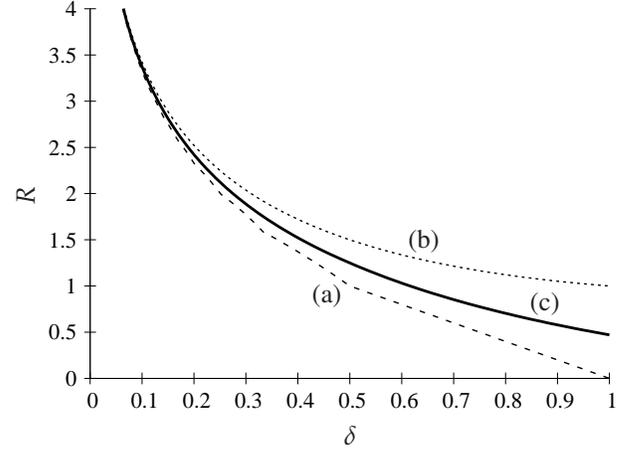}
  \caption{
    Upper bounds on the size of error-correcting codes over
    permutations with the infinity norm (rate $R$ as a function
    of the normalized minimum distance $\delta$):
    (a) The code-anticode bound of \cite{TamSch10};
    (b) The ball-packing bound of \cite{TamSch10};
    (c) The ball-packing bound using the new bounds of this paper.
  }
  \label{fig:ecc}
\end{figure}

Previous works on error-correcting codes over permutations with the
infinity norm \cite{KloLinTsaTze10,TamSch10,YehSch16} used bounds on
the size of balls in this metric to state ball-packing and
Gilbert--Varshamov-like bounds. Since in this paper we improved the
lower bound on the size of balls, these new bounds affect the
ball-packing bound (stated in \cite{TamSch10}). If we consider
error-correcting codes in $S_n$ of rate $R$ and normalized distance
$\delta$, the ball-packing bound\footnote{We comment that in \cite{TamSch10}, $\delta$ was defined in a slightly different manner, by $\delta\defeq d/n$, where $d$ is the minimum distance of the code. We instead use $\delta\defeq d/(n-1)$, to be consistent with the normalization used throughout the paper. The change due to the difference in normalization is subsumed in the $o(1)$ additive factor in the upper bound on the rate.} states that
\begin{align*}
  2^{Rn} 
    &\leq 
       \frac{n!}{\abs{B_{\floorenv{(\delta (n-1) - 1)/2},n}}},
\end{align*}
or in asymptotic form,
\begin{align*}
  R 
    &\leq 
       \log_2 n 
       - 
       \log_2 e 
       - 
       \frac{1}{n}\log_2\abs{B_{\delta/2,n}} 
       + 
       o(1).
\end{align*}
Using the bounds on $\abs{B_{\delta/2,n}}$ known at that time, it was stated
in \cite[Th.~27]{TamSch10} that
\begin{align*}
  R
    &\leq 
       \delta 
       + 
       \log_2 \frac{1}{\delta} 
       + 
       o(1).
\end{align*}
However, now we can use the improved lower bounds $\bl_2$ and $\bl_3$,
and obtain a stronger asymptotic form for the ball-packing bound,
\begin{align*}
  R 
    &\leq 
       \begin{cases}
         \frac{\delta}{2} + \log_2 \frac{1}{\delta} + o(1) 
           & 0 \leq \frac{\delta}{2} \leq \xi, \\
         (\log_2(e)-1)(\delta-1) + \log_2 \frac{1}{\delta} + 1 \\
         \quad - \log_2(\log_2(e)) + o(1)
           & \xi\leq \frac{\delta}{2} \leq \frac{1}{2},
       \end{cases}
\end{align*}
where $\xi \defeq
\frac{2-\log_2(e)-\log_2(\log_2(e))}{3-2\log_2(e)}\approx 0.249$ is
the crossover point of curves (b) and (c) in
Fig.~\ref{fig:bounds}. The ball-packing bound for error-correcting
codes is shown in Fig.~\ref{fig:ecc}, before and after the improvement
of this paper. We do note that while the improvement in the bound is
substantial, it is still weaker than the code-anticode bound described
in \cite{TamSch10}. (However, the bound in \cite{TamSch10} does not
have the geometric interpretation of packing balls.)

In contrast, in the case of covering codes over permutations with the
infinity norm, the new bounds in this paper do improve the best upper
bounds on the rate of the codes. Let us now consider covering codes in
$S_n$ of rate $R$ and normalized covering radius $\rho$. The upper
bound of \cite[Th.~3]{FarSchBru16a} is
\begin{align*}
  2^{Rn} 
    &\leq 
       \frac{n!(1+\ln(n!))}
            {\abs{B_{\rho,n}}}.
\end{align*}
Using the asymptotic bounds on $\abs{B_{\rho,n}}$ known at that time,
an asymptotic form was given in \cite[Th.~15]{FarSchBru16a} as
\begin{align*}
  R 
    &\leq 
       \begin{cases}
         2\rho + \log_2\frac{1}{\rho} + o(1) 
           & 0\leq \rho\leq \frac{1}{2},\\
         2(1-\rho) + o(1) 
           & \frac{1}{2}\leq \rho \leq 1.
       \end{cases}
\end{align*}
Now, using $\bl_2$ and $\bl_3$, we can state an improved upper bound
\begin{align*}
  R 
    &\leq 
       \begin{cases}
         \rho - 1 + \log_2\frac{1}{\rho} + o(1)  
           & 0\leq \rho\leq \xi, \\
         (2\rho-1)(\log_2(e)-1)\\
         \quad +\log_2\frac{1}{\rho}+\log_2(\log_2(e)) 
           & \xi \leq \rho\leq \frac{1}{2}, \\
         \log_2(\that)-\log_2(\log_2(e))\\
         \quad -(2\rho-1)\that-\log_2(1-\rho) 
           & \frac{1}{2}\leq \rho\leq 1,
       \end{cases}
\end{align*}
where $\xi$ as defined above, and $\that$ as defined in
Theorem~\ref{th:results:newgap:1:3}.

\begin{figure}
  \psfrag{xex}{\parbox{1cm}{\vskip0.4cm (a)}}
  \psfrag{xdx}{\parbox{2cm}{\vskip0.4cm \hskip0.9cm (b)}}
  \psfrag{xfx}{\parbox{1cm}{\hskip0.1cm (c)}}
  \psfrag{delta}{\parbox{2cm}{\hskip0.2cm $\rho$}}
  \psfrag{rrr}{\parbox{1cm}{\vskip0.7cm $R$}}
  \centering
  \includegraphics[scale=0.7]{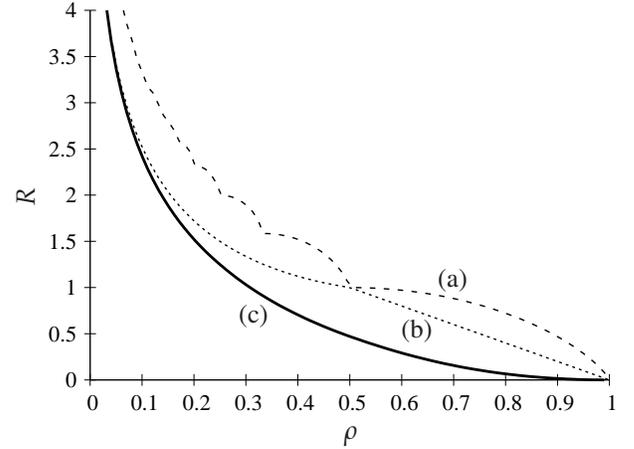}
  \caption{
    Upper bounds on the size of optimal covering codes over
    permutations with the infinity norm (rate $R$ as a function
    of the normalized covering radius $\rho$):
    (a) The covering-code construction of \cite{FarSchBru16a};
    (b) The upper bound of \cite{FarSchBru16a};
    (c) The upper bound using the new bounds of this paper.
  }
  \label{fig:cover}
\end{figure}

We observe that the largest improvement in Fig.~\ref{fig:bounds}
(between the previous lower bounds and our new lower bounds) occurs at
$\rho=\frac{1}{2}$. This manifests in Fig.~\ref{fig:cover} in a
similar manner, showing the largest improvement between curves (b) and
(c) occurring at $\rho=\frac{1}{2}$. However, in Fig.~\ref{fig:ecc},
the largest improvement between curves (b) and (c) occurs at
$\delta=1$. This is due to the fact that the ball-packing bound uses
$\abs{B_{\delta/2,n}}$.

\subsection{Open Problems}
\label{sec:open:problems:1}

We now turn to discuss some open problems. As mentioned at the
beginning of Section~\ref{sec:second:class:Q:matrices:1}, the $Q$
matrix that we use there for the case $\frac{1}{2} < \rho < 1$
maximizes the right-hand side of~\eqref{eq:th:sinkhornperlower} for $M
= A_{r,n}$. This is in contrast to the $Q$ matrix that we use for the
case $0 < \rho \leq 1/2$, which in general does not maximize the
right-hand side of~\eqref{eq:th:sinkhornperlower} for $M =
A_{r,n}$. We leave it as an open problem to find the $Q$ matrix that
maximizes the right-hand side of~\eqref{eq:th:sinkhornperlower} for
the case $0 < \rho \leq 1/2$. (So far, analytical considerations,
along with some numerical evidence for somewhat small choices of $n$
and $r$, have not led to a closed-form expression for the optimal $Q$
matrix.)

We mention another open problem. Instead of
Theorem~\ref{th:sinkhornperlower}, one can also use the following
approach to obtain a lower bound on $\log_2 \per(M)$. Namely, let
$\per_{\mathrm{B}}(M)$ be the Bethe permanent of an $n \times
n$-matrix with non-negative entries~\cite{Von13}. The following
theorem is due to Gurvits~\cite{Gur11}. (See also the discussion
in~\cite{Von13}.)

  \begin{theorem}
    \label{th:betheperlower}
    Let $M \defeq (m_{i,j})$ be an $n \times n$ matrix with
    non-negative entries and $\per(M) > 0$. Let $Q_{r,n} \defeq
    (q_{i,j})$ be any $n \times n$ doubly-stochastic matrix such that
    $q_{i,j}=0$ whenever $m_{i,j}=0$. Then
    \begin{align*}
      &\log_2 \per(M) \\
      &\quad 
       \geq
         \log_2 \per_{\mathrm{B}}(M) \\
      &\quad
       \geq \!
         \sum_{i,j \in [n]} \!
           \parenv{\!
                   -
                   q_{i,j}\log_2 \! \left( \frac{q_{i,j}}{m_{i,j}} \right)
                   +
                   (1 \! - \! q_{i,j}) \log_2(1 \! - \! q_{i,j})
                   \!
                  } \! .
    \end{align*}
    \theoremend
  \end{theorem}

For the asymptotic setup of interest in this paper, \ie, $r \defeq
\rho \cdot (n \! - \! 1)$, and the $Q$-matrices which were defined in
Sections~\ref{sec:first:class:Q:matrices:1}
and~\ref{sec:second:class:Q:matrices:1}, it turns out that
Theorems~\ref{th:sinkhornperlower} and~\ref{th:betheperlower} lead to
the same lower bounds (modulo $o(n)$ terms) except for the boundary
case $\rho = 0$. It is conceivable that an optimal choice of a
$Q$-matrix for Theorem~\ref{th:betheperlower} may result in a lower
bound that outperforms the bounds obtained in this paper. We leave it
as an open problem to analytically find the $Q$ matrix which maximizes
the right-hand side of the expression in
Theorem~\ref{th:betheperlower} for $M = A_{r,n}$.

\section*{Acknowledgments}

The authors would like to thank the associate editor and the anonymous
reviewers, whose comments helped to improve the presentation of this paper.

\appendices

\section{Proof of Theorem~\ref{th:known:bounds:1}}
\label{sec:proof:th:known:bounds:1}

To the best of our knowledge, the tightest known bounds for balls in $(S_n,
d_\infty)$ are as follows:
\begin{align*}
  \bl_1(\rho,n)
    &\leq \abs{B_{\rho,n}} \leq \bu_1(\rho,n),
\end{align*}
where
\begin{align*}
  \bl_1(\rho,n) 
    &\defeq
       \begin{cases}
         \frac{n! \cdot (2r+1)^n}{2^{2r} \cdot n^n} 
           & 0 < \rho \leq \frac{1}{2} \\[0.15cm]
         \frac{n! \cdot (2r+1)^n}{2^{n} \cdot n^n} 
           & \frac{1}{2} \leq \rho < 1
       \end{cases} \ , \\
  \bu_1(\rho,n)
    &\defeq
       \begin{cases}
         \big( (2r+1)! \big)^{\frac{n-2r}{2r+1}}
         \prod_{i=r+1}^{2r}(i!)^{\frac{2}{i}}
           &\!\!\! 0 < \rho \leq \frac{1}{2} \\[0.15cm]
         \parenv{n!}^{\frac{2r+2-n}{n}}
         \prod_{i=r+1}^{n-1}(i!)^{\frac{2}{i}}
           &\!\!\! \frac{1}{2} \leq \rho < 1
       \end{cases}
\end{align*}
where $r \defeq \rho \cdot (n \! - \! 1)$. These bounds are a consequence of
the following results:
\begin{itemize}

\item For the range $0 < \rho\leq \frac{1}{2}$, the upper bound
  was given in~\cite[Eq. (4)]{Klo08}, and the lower bound
  was given in~\cite[Eq. (3) and (5)]{Klo11}.

\item For the range $\frac{1}{2}\leq \rho <1$, the upper bound was given
  in~\cite[Lemma 25]{TamSch10}.

\item For the range $\frac{1}{2}\leq \rho <1$, a slightly weaker lower
  bound was given in~\cite[Lemma 14]{FarSchBru16a}. However, the bound
  we cite here, though never presented explicitly, may be deduced
  from~\cite[Eq. (3) and (5)]{Klo11} while noting that (in the
  notation of \cite{Klo11})
  \begin{align*}
  \per\big( A^{(r,n)} \big) 
    &= \frac{1}{2^n} 
         \cdot
         \per\big( 2 A^{(r,n)} \big) \\
    &\geq 
       \frac{1}{2^n}
         \cdot
         \per\big( B^{(r,n)} \big) \\
    &\geq 
       \frac{n! \cdot (2r+1)^n}
            {2^n\cdot n^n}.
  \end{align*}
\end{itemize}

We would now like to convert these bounds to a more pleasing
asymptotic form. We start with the range $0 < \rho\leq
\frac{1}{2}$. For $\bl_1(\rho,n)$ we obtain
\begin{align*}
  \bl_1(\rho, n)
    &= \frac{n! \cdot (2\rho \cdot (n\!-\!1) + 1)^n}{2^{2\rho \cdot (n\!-\!1)}n^n}
     = \frac{(2\rho n)^n}{2^{2\rho n}e^n}\cdot 2^{o(n)},
\end{align*}
where we used Stirling's approximation.

On the other hand, the expression for $\bu_1(\rho,n)$ follows from
\begin{align*}
  \bu_1(\rho, n)
    &= \big( 
         (2\rho (n \! - \! 1)+1)! 
       \big)^{\frac{n-2\rho \cdot (n\!-\!1)}{2\rho \cdot (n\!-\!1)+1}}
         \ \cdot
         \prod_{i=\rho \cdot (n\!-\!1)+1}^{2\rho \cdot (n\!-\!1)}
           (i!)^{\frac{2}{i}} \\
    &= \parenv{\frac{2\rho n}{e}}^{n(1-2\rho)}
         \cdot
         \parenv{\prod_{i=\rho \cdot (n\!-\!1)+1}^{2\rho \cdot (n\!-\!1)} 
                   (i!)^{\frac{2}{i}}
                      }
         \cdot 
         2^{o(n)} \\
    &= \parenv{\frac{2\rho n}{e}}^{n(1-2\rho)}\\
    &\quad\ \cdot
               \parenv{\prod_{i=\rho \cdot (n\!-\!1)+1}^{2\rho \cdot (n\!-\!1)}
                 \parenv{
                   \parenv{\frac{i}{e}}^2 2^{o(1)}}
                 }
                 \cdot 
                 2^{o(n)} \\
    &= \parenv{\frac{2\rho n}{e}}^{n(1-2\rho)}\\
    &\quad\  \cdot
               \frac{1}{e^{2\rho \cdot (n\!-\!1)}}\cdot
       \parenv{\frac{(2\rho \cdot (n\!-\!1))!}
              {(\rho \cdot (n\!-\!1))!}}^2 \cdot 2^{o(n)}\\
    &= \parenv{\frac{2\rho n}{e}}^{n(1-2\rho)}\cdot
               \frac{1}{e^{2\rho n}}\cdot
       2^{4\rho n} \cdot \parenv{\frac{\rho n}{e}}^{2\rho n} \cdot 2^{o(n)}\\
    &= \parenv{\frac{2\rho n}{e}}^{n}
         \cdot
         \parenv{\frac{2}{e}}^{2\rho n}
         \cdot
         2^{o(n)},
\end{align*}
where we used Stirling's approximation whenever a factorial appears.
The case of $\frac{1}{2}\leq \rho < 1$ is handled analogously. Namely,
after following similar steps we get
\begin{align*}
  \bl_1(\rho,n)
  &= \frac{(2\rho n)^n}{2^{n}e^n}\cdot 2^{o(n)},\\
  \bu_1(\rho,n)
    &= \frac{n^n}{e^{n(3-2\rho)}\rho^{2\rho n}}
         \cdot
         2^{o(n)}.
\end{align*}
Clearly, the obtained expressions for $\bu_1$ and $\bl_1$ imply the
expressions given in the theorem statement.

\section{Proof of Theorem~\ref{th:known:bounds:1:mod}}
\label{sec:proof:th:known:bounds:1:mod}

We recall that the asymptotic regime of interest to us is $r = \rho \cdot (n
\! - \! 1)$ for some fixed $\rho$. The conjecture of \cite{Klo11}, proven in
\cite{GuoYan16}, asserts that for $0 < \rho \leq \frac{1}{2}$,
\begin{align}
  \label{eq:conj1}
  \abs{B_{\rho,n}} 
    &> \frac{\sqrt{2\pi(n+2r)}}
            {\omega_r^2}
       \cdot
       \parenv{\frac{2r+1}{e}}^n,
\end{align}
where
\begin{align}
  \label{eq:conj2}
  \omega_r 
    &\defeq 
       \frac{\Omega_r \cdot e^r}{(2r+1)^r} \ ,
\end{align}
and
\begin{align*}
  \Omega_r
    &\defeq
       \sum_{m=0}^r 
         \binom{r}{m}
         \cdot
         (m+1)^r \ .
\end{align*}

In order to complete the analysis, we need to find an asymptotic expression
for $\Omega_r$. To that end, let us denote
\begin{align*}
  \tilde \Omega_r
    &\defeq
       \max
         \left\{
             \binom{r}{m} \cdot (m+1)^r 
           \ \middle| \ 
             0 \leq m \leq r
         \right\}.
\end{align*}
Then
\begin{align*}
  \tilde \Omega_r
    &\leq 
       \Omega_r 
     \leq 
       (r+1) \cdot \tilde \Omega_r.
\end{align*}
In the limit $n \to \infty$, these inequalities imply
\begin{align*}
  \log_2(\Omega_r)
    &= \log_2(\tilde \Omega_r) + o(n).
\end{align*}

In order to find $\tilde \Omega_r$, we look for the maximal summand in the
definition of $\Omega_r$. We do so by looking at the ratio of two successive
summands,
\begin{align*}
  \frac{\binom{r}{m} \cdot (m+1)^r}
       {\binom{r}{m-1} \cdot m^r},
\end{align*}
and note that this ratio is monotone decreasing in $m$. We denote $m =
\mu \cdot r$ for some suitable $\mu$, and thus we would like to find
the value of $\mu$ for which
\begin{align*}
  \frac{\binom{r}{\mu r} \cdot (\mu r+1)^r}
       {\binom{r}{\mu r-1} \cdot (\mu r)^r}
    &= 1.
\end{align*}
Since $r$ tends to infinity, in the limit, this equation becomes
\begin{align*}
  \frac{1-\mu}{\mu}
    \cdot
    e^{1/\mu}
    &= 1,
\end{align*}
whose exact solution is
\begin{align*}
  \mu
    &= \frac{1}{1+W(e^{-1})}
     \defeq \mu^*.
\end{align*}
Thus the value of $m$ we are looking for is
\begin{align*}
  m 
    &= \mu^* \cdot r \cdot (1+o(1)),
\end{align*}
and so
\begin{align*}
  \log_2(\Omega_r)
    &= \log_2(\tilde \Omega_r) + o(n)\\
    &= n
         \cdot
         \big(
           h(\mu^*) 
           + 
           \log_2(\mu^*\rho n)
         \big)
         \cdot
         \rho
         +o(n).
\end{align*}
Plugging this back into~\eqref{eq:conj1} and~\eqref{eq:conj2}, we
obtain the promised result.

\section{Proof of Lemma~\ref{lemma:firstQ}}
\label{sec:proof:lemma:firstQ}

We begin by noting that the support of $Q_{r,n}$ is the same as that
of $A_{r,n}$. We need to verify that all entries of $Q_{r,n}$ are
non-negative, that all columns sum to~$1$, and that all rows sum
to~$1$. It follows immediately from the definition of $Q_{r,n}$ that
all entries are non-negative. Because $Q_{r,n}$ is symmetric, it only
remains to show that all columns sum to~$1$.

First let us consider the case of $0\leq r\leq \frac{n-1}{2}$. In this case,
for columns $1 \leq j \leq r$ we have
\begin{align*}
  \sum_{i=1}^{n} 
    q_{i,j} 
    &= \sum_{i=1}^{r+1-j} 
         \frac{2}{2r+1} 
       + 
       \sum_{i=r+2-j}^{r+j} 
         \frac{1}{2r+1} 
     = 1.
\end{align*}
A similar statement holds for columns $n - r + 1 \leq j \leq n$. For the rest
of the columns, \ie, for $r + 1 \leq j \leq n - r$, we have
\begin{align*}
  \sum_{i=1}^n 
    q_{i,j}
    &= \sum_{i=j-r}^{j+r} 
         \frac{1}{2r+1} 
     = 1.
\end{align*}

Let us now consider the case $\frac{n-1}{2} \leq r \leq n - 1$. For
  columns $1 \leq j\leq n-r-1$ we have
\begin{align*}
  \sum_{i=1}^n 
    q_{i,j} 
    &= \sum_{i=1}^{n-r-j} 
         \frac{2}{n} 
       + 
       \sum_{i=n-r-j+1}^{r+j} 
         \frac{1}{n} 
     = 1,
\end{align*}
and a symmetric claim holds for columns $r + 2 \leq j \leq n$. Finally, for
columns $n - r \leq j \leq r + 1$ we have
\begin{align*}
  \sum_{i=1}^n 
    q_{i,j} 
    &= \sum_{i=1}^n 
         \frac{1}{n}
     = 1.
\end{align*}

\section{Proof of Theorem~\ref{th:newlow:2}}
\label{sec:proof:th:newlow:2}

Let $r \defeq \rho \cdot (n \! - \! 1)$. We distinguish two cases, namely the
case $0 < \rho \leq \frac{1}{2}$ and the case $\frac{1}{2} \leq \rho < 1$.

Consider the first case, \ie, $0 < \rho \leq \frac{1}{2}$. We make the
following observations about the matrix $Q_{r,n}$ in
Definition~\ref{def:first:class:Q:matrices}:
\begin{itemize}

\item $r (r+1)$ entries take on the value $\frac{2}{2r+1}$,

\item $n(2r+1)-2r(r+1)$ entries take on the value $\frac{1}{2r+1}$,

\item the remaining entries take on the value $0$.

\end{itemize}
We obtain
\begin{align*}
  \log_2 & \abs{B_{\rho,n}} \\
    &= \log_2 \per(A_{\rho,n}) \\
    &\geq
       \log_2 \!
         \left(
           \frac{n!}{n^n}
         \right)
       -
       r \cdot (r+1) \cdot \frac{2}{2r+1}
         \cdot
         \log_2 \!
         \left(
           \frac{2}{2r+1}
         \right) \\
    &\quad\
       -
       \big( n(2r+1)-2r(r+1) \big)
         \cdot
         \frac{1}{2r+1}
         \cdot
         \log_2
           \left(
             \frac{1}{2r+1}
           \right) \\
  &= \log_2(n!)
        -
        \frac{2r \cdot (r+1)}{2r+1}
        +
        n
          \cdot
          \log_2 \!
            \left(
              \frac{2r+1}{n}
            \right),
\end{align*}
where the first equality follows from
Lemma~\ref{lemma:relationship:permanent:ball:size:1} and where the inequality
follows from Theorem~\ref{th:sinkhornperlower} with $M \defeq A_{\rho,n}$ and
with $Q \defeq Q_{r,n}$, where $Q_{r,n}$ was specified in
Definition~\ref{def:first:class:Q:matrices}. For the asymptotics, we
make note of the following:
\begin{align*}
  \log_2(n!) & = \log_2\parenv{\parenv{\frac{n}{e}}^n\cdot 2^{o(n)}} \\
  & = n\log_2(n) - n\log_2(e) + o(n),
\end{align*}
by Stirling's approximation, as well as
\begin{align*}
  \frac{2r\cdot(r+1)}{2r+1} & = \frac{2\rho(n-1)(\rho(n-1)+1)}{2\rho(n-1)+1} \\
  &= \rho n + o(n),
\end{align*}
and
\begin{align*}
  n\cdot\log_2\parenv{\frac{2r+1}{n}} &= n\cdot\log_2\parenv{\frac{2\rho(n-1)+1}{n}} \\
  &= n\cdot\log_2(2\rho + o(1)) \\
  &= n\cdot\log_2(2\rho)+o(n).
\end{align*}
Combining these together, asymptotically we get
\begin{align*}
  n\cdot\log_2(n)
  -
  n
    \cdot
    \big[
      \log_2(e)
      -
      1
      +
      \rho
      -
      \log_2(\rho)
    \big]
  +
  o(n),
\end{align*}
which confirms the expression in the theorem statement.

We now turn to the second case, \ie, $\frac{1}{2} \leq \rho < 1$. We
make the following observations about the matrix $Q_{r,n}$ in
Definition~\ref{def:first:class:Q:matrices}:
\begin{itemize}

\item $(n-r-1)(n-r)$ entries take on the value $\frac{2}{n}$,

\item $n^2-2(n-r-1)(n-r)$ entries take on the value $\frac{1}{n}$,

\item the remaining entries take on the value $0$.

\end{itemize}
We therefore obtain
\begin{align*}
  \log_2 & \abs{B_{\rho,n}} \\
    &= \log_2 \per(A_{\rho,n}) \\
    &\geq
       \log_2 \!
         \left(
           \frac{n!}{n^n}
         \right)
       -
       (n-r-1)(n-r)
         \cdot
         \frac{2}{n}
         \cdot
         \log_2 \!
           \left(
             \frac{2}{n}
           \right) \\
    &\quad\
       -
       (n^2-2(n-r-1)(n-r))
         \cdot
         \frac{1}{n}
         \cdot
         \log_2 \!
           \left(
             \frac{1}{n}
           \right) \\
    &= \log_2(n!)
       -
       \frac{2 \cdot (n-r-1) \cdot (n-r)}{n},
\end{align*}
where the first equality follows from
Lemma~\ref{lemma:relationship:permanent:ball:size:1} and where the
inequality follows from Theorem~\ref{th:sinkhornperlower} with $M
\defeq A_{\rho,n}$ and with $Q \defeq Q_{r,n}$, where $Q_{r,n}$ was
specified in Definition~\ref{def:first:class:Q:matrices}. For an
asymptotic expression we follow the same steps as in the previous
case. Thus, asymptotically, the last expression becomes
\begin{align*}
  n\cdot\log_2(n)
  -
  n\cdot\log_2(e)
  -
  2 n 
    \cdot
    (1-\rho)^2
  +
  o(n),
\end{align*}
which confirms the expression in the theorem statement.

\section{Proof of Lemma~\ref{lemma:newlow:3:1:Q:properties}}
\label{sec:proof:lemma:newlow:3:1:Q:properties}

The following lemma collects some results that will prove useful for
simplifying some upcoming computations in this appendix and also in
Appendix~\ref{sec:proof:th:newlow:3:1}.

\begin{lemma}
  \label{lemma:newlow:3:1:useful:sums}
  We define
  \begin{alignat*}{2}
    S_r^{(0)}
      &\defeq
         \sum_{\ell=0}^{r}
           \alpha_r^{\ell}
     &&= \frac{\alpha_r^{r+1} - 1}
              {\alpha_r - 1} \\
    &&&= \frac{\alpha_r}
              {\alpha_r - 1} \ , \\
    S_r^{(1)}
      &\defeq
         \sum_{\ell=0}^{r}
           \ell \cdot \alpha_r^{\ell}
     &&= \frac{r \cdot \alpha_r^{r+2} - (r+1) \cdot \alpha_r^{r+1} + \alpha_r}
              {(\alpha_r - 1)^2} \\
    &&&= \frac{r \cdot \alpha_r^2 - r - 1}
              {(\alpha_r - 1)^2} \ , \\
    S_r^{(2)}
      &\defeq
         \sum_{\ell=0}^{r}
           \ell^2 \cdot \alpha_r^{\ell}
     &&= \frac{r^2 \cdot \alpha_r^{r+1} - 2 \cdot S_{r}^{(1)}}
              {\alpha_r - 1}
         +
         \frac{\alpha_r^{r+1} - \alpha_r}
              {(\alpha_r - 1)^2} \\
    &&&= \frac{r^2 \cdot (\alpha_r + 1)}
              {\alpha_r - 1}
         +
         \frac{1}
              {(\alpha_r - 1)^2} \\
    &&&\quad\ 
         -
         2
         \cdot
         \frac{r \cdot \alpha_r^2 - r - 1}
              {(\alpha_r - 1)^3} \ .
  \end{alignat*}
\end{lemma}

\begin{IEEEproof}
  In each case, the the first summation expression is obtained by standard
  algebraic techniques, whereas the second summation expression is obtained by
  simplifying the first summation expression with the help
  of~\eqref{eq:def:newlow:3:1:alpha}.
\end{IEEEproof}

We now continue with the proof of
Lemma~\ref{lemma:newlow:3:1:Q:properties}. First we note that by
definition, $Q_{r,n}$ has the same support as $A_{r,n}$. We need to
verify that all entries of $Q_{r,n}$ are non-negative, that all
columns sum to~$1$, and that all rows sum to~$1$. It follows
immediately from the definition of $Q_{r,n}$ that all entries are
non-negative. Because $Q_{r,n}$ is symmetric, it only remains to show
that all columns sum to~$1$.

For $1 \leq j \leq r+1$, we obtain
\begin{align*}
  \sum_{i=1}^{n}
    q_{i,j}
    &= C
         \cdot
         \sum_{i=1}^{r+1}
           \alpha_r^{(r+1-i)+(r+1-j)}
       +
       C
         \cdot
         \sum_{i=r+2}^{r+j}
           \alpha_r^{i-j} \\
    &= C
         \cdot
         \alpha_r^{r+1-j}
         \cdot
         S_r^{(0)}
       +
       C
         \cdot
         \frac{\alpha_r^{r+1} - \alpha_r^{r+2-j}}
              {\alpha_r - 1} \\
    &= C
         \cdot
         \alpha_r^{r+1-j}
         \cdot
         \frac{\alpha_r}{\alpha_r-1}
       +
       C
         \cdot
         \frac{\alpha_r^{r+1} - \alpha_r^{r+2-j}}
              {\alpha_r - 1} \\
    &= 1.
\end{align*}

For $n \! - \! r \leq j \leq n$, because of symmetries of the $Q_{r,n}$ matrix,
the calculations are analogous to the calculations for $1 \leq j \leq r+1$.

Finally, for $r + 2 \leq j \leq n \! - \! r \! - \! 1$, we obtain
\begin{align*}
  \sum_{i=1}^{n}
    q_{i,j}
    &= C
         \cdot
         \sum_{i=j-r}^{j+r}
           \alpha_r^{|i-j|}
     = 2 \cdot C \cdot S_r^{(0)} - C \cdot \alpha_r^0
     = 1.
\end{align*}

\section{Proof of Theorem~\ref{th:newlow:3:1}}
\label{sec:proof:th:newlow:3:1}

Recall that $r \defeq \rho \cdot (n \! - \! 1)$. In the following, in order to
simplify the notation, we define $\alpha \defeq \alpha_r$. As in
Appendix~\ref{sec:proof:th:newlow:2}, the proof here is based on
Lemma~\ref{lemma:relationship:permanent:ball:size:1} and
Theorem~\ref{th:sinkhornperlower}. To this end, we compute the quantity
\begin{align*}
  T
    &\defeq
       \sum_{i,j \in [n]}
         q_{i,j}\log_2 \frac{q_{i,j}}{m_{i,j}} \ ,
\end{align*}
where $Q_{r,n} = (q_{i,j})$ is the matrix specified in
Definition~\ref{def:newlow:3:1:matr:Q} and where $M = (m_{i,j}) =
A_{r,n}$. We decompose $T$ as follows
\begin{align*}
  T
    &= T_1 + T_2 + T_3 + T_4 + T_5 \ ,
\end{align*}
where
\begin{align*}
  T_1
    &\defeq
       \sum_{j=1}^{r+1} \ 
         \sum_{i=1}^{r+1}
           q_{i,j}\log_2 \frac{q_{i,j}}{m_{i,j}} \ , \\
  T_2
    &\defeq
       \sum_{j=2}^{r+1} \ 
         \sum_{i=r+2}^{j+r}
           q_{i,j}\log_2 \frac{q_{i,j}}{m_{i,j}} \ , \\
  T_3
    &\defeq
       \sum_{j=r+2}^{n-r-1} \ 
         \sum_{i=j-r}^{j+r}
           q_{i,j}\log_2 \frac{q_{i,j}}{m_{i,j}} \ , \\
  T_4
    &\defeq
       \sum_{j=n-r}^{n-1} \ 
         \sum_{i=j-r}^{n-r-1}
           q_{i,j}\log_2 \frac{q_{i,j}}{m_{i,j}} \ , \\
  T_5
    &\defeq
       \sum_{j=n-r}^{n} \ 
         \sum_{i=n-r}^{n}
           q_{i,j}\log_2 \frac{q_{i,j}}{m_{i,j}} \ .
\end{align*}
Because of the symmetries of the setup, we have $T_1 = T_5$ and $T_2 =
T_4$. Therefore, we only need to determine $T_1$, $T_2$, and $T_3$. We get,
using the notation from Lemma~\ref{lemma:newlow:3:1:useful:sums} in
Appendix~\ref{sec:proof:lemma:newlow:3:1:Q:properties},
\begin{align*}
  T_1
    &= \sum_{j=1}^{r+1} \ 
         \sum_{i=1}^{r+1}
           C
           \cdot
           \alpha^{(r+1-i)+(r+1-j)}
           \log_2 \!
             \left(
               C
               \cdot
               \alpha^{(r+1-i)+(r+1-j)}
             \right) \\
    &= C
         \cdot
         \big( S_r^{(0)} \big)^2
         \cdot
         \log_2(C)
       +
       2
         \cdot
         C
         \cdot
         S_r^{(0)}
         \cdot
         S_r^{(1)}
         \cdot
         \log_2(\alpha) \ , \\
  T_2
    &= \sum_{j=2}^{r+1} \ 
         \sum_{i=r+2}^{j+r}
           q_{i,j}\log_2 \frac{q_{i,j}}{m_{i,j}} \\
    &= \sum_{\ell=0}^{r}
         \ell
         \cdot
         C
         \cdot
         \alpha^{\ell}
         \cdot
         \log_2 \!
           \left(
             C
             \cdot
             \alpha^{\ell}
           \right) \\
    &= C
         \cdot
         S_r^{(1)}
         \cdot
         \log(C)
       +
       C
         \cdot
         S_r^{(2)}
         \cdot
         \log(\alpha) \ , \\
  T_3
    &= \sum_{j=r+2}^{n-r-1} \ 
         \sum_{i=j-r}^{j+r}
           q_{i,j}\log_2 \frac{q_{i,j}}{m_{i,j}} \\
    &= \big(
         (n \! - \! r \! - \! 1)
         -
         (r \! + \! 2)
         +
         1
       \big)
       \cdot
       \sum_{\ell=-r}^{r}
         C
         \cdot
         \alpha^{|\ell|}
         \cdot
         \log_2
           \left(
             C
             \cdot
             \alpha^{|\ell|}
           \right) \\
   &= (n \! - \! 2r \! - \! 2)
        \cdot
        C
        \cdot
        \big(
          2 S_r^{(0)}
          -
          1
        \big)
        \cdot
        \log(C) \\
    &\quad\ 
       +
       (n \! - \! 2r \! - \! 2)
       \cdot
       C
       \cdot
       \big(
         2 S_r^{(1)}
         -
         0
       \big)
       \cdot
       \log(\alpha) \ .
\end{align*}

Recall that $\alpha$ satisfies $\alpha^{r+1} - \alpha - 1 = 0$. The following
lemma gives an approximation of $\alpha$ which is precise enough for the
upcoming computations.

\begin{lemma}
  It holds that
  \begin{align*}
    \alpha 
      &= 1 + \frac{\ln(2)}{r} + o(1/r).
  \end{align*}
\end{lemma}

\begin{IEEEproof}
  Consider the following two functions
  \begin{align*}
    \underline{g}(r)
      &\defeq
         \left.
           \underline{\alpha}^{r+1}
           -
           \underline{\alpha}
           -
           1
         \right|_{\underline{\alpha} = 1 + \frac{\ln(2)}{r+1}} \ , \\
    \overline{g}(r)
      &\defeq
         \left.
           \overline{\alpha}^{r+1}
           -
           \overline{\alpha}
           -
           1
         \right|_{\overline{\alpha} = 1 + \frac{\ln(2)}{r}} \ .
  \end{align*}
  One can show that $\underline{g}(r)$ is a strictly increasing function of
  $r > 0$, ultimately converging to $0$ as $r \to \infty$, and that
  $\overline{g}(r)$ is a strictly decreasing function of $r > 0$, ultimately
  converging to $0$ as $r \to \infty$. (We omit the straightforward, but
  tedious, details.)

  From these observations, it follows that $\alpha$ satisfies
  \begin{align*}
    1 + \frac{\ln(2)}{r+1}
      &=
         \underline{\alpha}
       \leq
         \alpha
       \leq
         \overline{\alpha}
       = 1 + \frac{\ln(2)}{r},
  \end{align*}
  which implies the expression in the lemma statement.
\end{IEEEproof}

In the following, we will therefore use $\alpha = 1 + \frac{\ln(2)}{r} +
o(1/r)$. With this, we obtain $C = \frac{\ln(2)}{2 \rho n} + o(1/n)$.

Putting everything together, and using
Lemma~\ref{lemma:relationship:permanent:ball:size:1} and
Theorem~\ref{th:sinkhornperlower}, we get
\begin{align*}
  &
  \log_2|B_{\rho,n}| \\
    &\geq
       \log_2 \!
           \left(
             \frac{n!}{n^n}
           \right)
       -
       T \\
    &= n \cdot \log_2(n) \\
    &\quad\ 
       -
       n
       \cdot 
         \Big[
           \big( \log_2(e) - 1 \big) \cdot 2\rho
           -
           \log_2(\rho)
           -
           \log_2\big( \log_2(e) \big)
           +
           1
         \Big] \\
     &\quad\  
       +
       o(n).
\end{align*}

\section{Proof of Lemma~\ref{lemma:newlow:3:2:matr:Q}}
\label{sec:proof:lemma:newlow:3:2:matr:Q}

We begin by noting that the support of $Q_{r,n}$ is the same as that
of $A_{r,n}$. We need to verify that all entries of $Q_{r,n}$ are
non-negative, that all columns sum to~$1$, and that all rows sum
to~$1$. It follows immediately from the definition of $Q_{r,n}$ that
all entries are non-negative. Because $Q_{r,n}$ is symmetric, it only
remains to show that all columns sum to~$1$. In order to simplify the
notation, in the following we set $\alpha \defeq \alpha_{r,n}$.

For $1 \leq j \leq n-r$, we obtain
\begin{align*}
  \sum_{i=1}^{n}
    q_{i,j}
    &= C
       \cdot
       \alpha^{n-r-j}
       \cdot
       \Big(
         \alpha^{n-r-1} + \cdots + \alpha^2 + \alpha + 1 \\
    &\hskip2.7cm
         + (2r - n) \cdot 1 \\ 
    &\hskip2.7cm
         + 1 + \alpha + \alpha^2 + \cdots + \alpha^{j-1} 
       \Big) \\
    &= C
       \cdot
       \alpha^{n-r-j}
       \cdot
       \bigg(
         \frac{\alpha^{n-r} - 1}{\alpha - 1}
         + 2r - n 
         + \frac{\alpha^j - 1}{\alpha - 1}
       \bigg) \\
    &= \alpha^{-j}
       \cdot
       \bigg(
         \alpha^{n-r}
         -
         1
         + (2r - n) \cdot (\alpha \! - \! 1)
         + \alpha^j
         -
         1
       \bigg) \\
    &= 1,
\end{align*}
where the third equality follows from plugging in the expression for $C$
from~\eqref{eq:def:newlow:3:2:C:1} and where the fourth equality follow from
using~\eqref{eq:def:newlow:3:2:alpha} to simplify the expression.

For $n-r+1 \leq j \leq r$, we obtain
\begin{align*}
  \sum_{i=1}^{n}
    q_{i,j}
    &= \sum_{i=1}^{n}
         q_{i,n-r}
     = 1,
\end{align*}
where the first equality follows from $q_{i,j} = q_{i,n-r}$, $i \in [n]$, and
where the second equality follows from the above computations.

For $r+1 \leq j \leq n$, we can use the symmetries of the matrix $Q_{r,n}$ and the
above computations to argue that $\sum_{i=1}^{n} q_{i,j} = 1$.

\section{Proof of Lemma~\ref{lemma:newlow:3:2:partial:result:1}}
\label{sec:proof:lemma:newlow:3:2:partial:result:1}

In the following, in order to simplify the notation, we define $\alpha \defeq
\alpha_{r,n}$. As in Appendices~\ref{sec:proof:th:newlow:2}
and~\ref{sec:proof:th:newlow:3:1}, the proof here is based on
Lemma~\ref{lemma:relationship:permanent:ball:size:1} and
Theorem~\ref{th:sinkhornperlower}. To this end, we compute the quantity
\begin{align*}
  T
    &\defeq
       \sum_{i,j \in [n]}
         q_{i,j}\log_2 \frac{q_{i,j}}{m_{i,j}} \ ,
\end{align*}
where $Q_{r,n}$ is the matrix specified in Definition~\ref{def:newlow:3:2:matr:Q}
and where $M = A_{r,n}$. We get
\begin{align*}
  T
    &= \sum_{i,j \in [n]}
         q_{i,j}
         \cdot
         \log_2
           \parenv{
             C \cdot \exp_2(\lambda_i) \cdot \exp_2(\lambda'_j)
           } \\
    &= n
         \cdot
         \log_2(C)
       +
       \sum_{i \in [n]}
         \lambda_i
       +
       \sum_{j \in [n]}
         \lambda'_j \\
    &= n
         \cdot
         \log_2(\alpha-1)
       -
       n
         \cdot
         (n-r)
         \cdot
         \log_2(\alpha) \\
     &\quad\ 
       +
       2
         \cdot
         (n-r)
         \cdot
         (n-r-1)
         \cdot
         \log_2(\alpha) \\
    &= n
         \cdot
         \log_2(\alpha-1)
       -
       (n-r)
       \cdot
         (2r-n+2)
         \cdot
         \log_2(\alpha).
\end{align*}
Using Lemma~\ref{lemma:relationship:permanent:ball:size:1} and
Theorem~\ref{th:sinkhornperlower}, we therefore obtain
\begin{align*}
  \log_2 & \abs{B_{\rho,n}} \\
    &= \log_2 \per(A_{\rho,n}) \\
    &\geq
       \log_2 \!
           \left(
             \frac{n!}{n^n}
           \right)
       -
       T \\
    &= \log_2(n!)
       -
       n
         \log_2(n)
       -
       n
         \cdot
         \log_2(\alpha-1) \\
    &\quad\ 
       +
       (n-r)
       \cdot
         (2r-n+2)
         \cdot
         \log_2(\alpha).
\end{align*}

\section{Proof of Lemma~\ref{lemma:newlow:3:2:alpha:simplification:1}}
\label{sec:proof:lemma:newlow:3:2:alpha:simplification:1}

Let $\alpha \defeq \alpha_{r,n}$. We define the function $f:\R\to\R$,
\begin{align*}
  f(x)
    &= x^{n-r}
       +
       (2r \! - \! n) \cdot x
       -
       (2r \! - \! n \! + \! 2).
\end{align*}
By definition, $\alpha$ is the unique positive root of $f(x)$. We note that
\begin{align*}
  f(1)
    &= 1 + (2r \! - \! n) - (2r \! - \! n \! + \! 2)
     = -1 
     < 0, \\
  f\bigl( 2^{\frac{1}{n-r}} \bigr)
    &= 2 + (2r \! - \! n) \cdot 2^{\frac{1}{n-r}} - (2r \! - \! n \! + \! 2)
     > 0.
\end{align*}
It follows that
\begin{align*}
  \alpha
    &\in 
       \left[ 1, 2^{\frac{1}{n-r}} \right].
\end{align*}
We rewrite $\alpha$ by introducing a real parameter $t \in [0,1]$,
\begin{align*}
  \alpha
    &= 1 + t \cdot \parenv{2^{\frac{1}{n-r}} \! - \! 1}.
\end{align*}
In order to find the value of $t$ we need to solve
\begin{align}
  0
    &= f(\alpha) \nonumber \\
    &= \parenv{1 + t \cdot \parenv{2^{\frac{1}{n-r}} \! - \! 1}}^{n-r}
       + (2r \! - \! n) \cdot t \cdot \parenv{2^{\frac{1}{n-r}} \! - \! 1}
       - 2. 
         \label{eq:troot}
\end{align}

Solving \eqref{eq:troot} is not easy. Instead of taking the direct route, we
observe that
\begin{align}
  \label{eq:thatroot}
  \lim_{n\to\infty}f(\alpha)
    &= 2^t
       +
       t \cdot \frac{(2\rho-1) \ln (2)}{1-\rho}
       -
       2.
\end{align}
We conveniently define the right-hand side of \eqref{eq:troot} as $g(t)$, and
the right-hand side of \eqref{eq:thatroot} as $\ghat(t)$. We would like to
find $t^*$ such that $g(t^*) = 0$, which appears to be a difficult
task. Instead, we find $\that$ such that $\ghat(\that) = 0$, and claim that it
is not too far from $t^*$.

We first note that $\that$ from \eqref{eq:defthat} indeed satisfies
$\ghat(\that)=0$. This is done by expanding $\ghat(\that)$ and
remembering that $e^{W(z)}=\frac{z}{W(z)}$.

We now need to bound $\abs{t^*-\that}$. Since $\ghat(t)$ is continuous and
monotone increasing in $[0,1]$,
\begin{align*}
  \abs{t^*-\that}
    &\leq
       \frac{\max_{x\in[0,1]}
               \abs{g(x)-\ghat(x)}}
            {\min_{x\in[0,1]}
               \abs{\frac{d}{dx}\ghat(x)}}.
\end{align*}
It is easy to verify that
\begin{align*}
  \abs{g(x)-\ghat(x)}
    &= \Theta\big( n^{-1} \big),
         \quad x \in [0,1],
\end{align*}
by noting that
\begin{align*}
  &
  \lim_{n\to\infty}
    n \cdot \parenv{g(x)-\ghat(x)} \\
    &= \frac{x\ln(2)}
            {2(1-\rho)^2} \\
       &\hskip0.85cm
        \cdot \big( (2\rho \! - \! 1) \cdot \ln(2)
                     -
                     2^x \cdot (x \! - \! 1) \cdot (1 \! - \! \rho) 
                       \cdot \ln(2)
                     - 2\rho
              \big).
\end{align*}
Furthermore, we get
\begin{align*}
  \min_{x\in[0,1]}
    \abs{\frac{d}{dx} \ghat(x)}
    &= \frac{\rho \cdot\ln(2)}
            {1-\rho}.
\end{align*}
Combining everything together, we get that the sought after~$t^*$, for which
$g(t^*) = 0$, is
\begin{align*}
  t^*
    &= \that
       +
       \Theta\big( n^{-1} \big)
\end{align*}
which completes the proof the lemma.

\section{Proof of Theorem~\ref{th:lambert:function:based:bound}}
\label{sec:proof:th:lambert:function:based:bound}

Let $\alpha \defeq \alpha_{r,n}$. With the help of
Theorem~\ref{th:sinkhornperlower} we obtain
\begin{align*}
  \log_2\abs{B_{r,n}}
    &\geq
       \log_2(n!)
       -
       n \cdot \log_2(n)
       -
       n \cdot \log_2(\alpha \! - \! 1) \\
    &\quad\ 
       +
       (n \! - \! r) \cdot (2r \! - \! n \! + \! 2) \cdot \log_2(\alpha) \\
    &= -
       n \cdot \log_2 (e)
       -
       n \cdot \log_2(\alpha \! - \! 1) \\
    &\quad\ 
       + (n \! - \! r) \cdot (2r \! - \! n \! + \! 2) \cdot \log_2(\alpha)
       + o(n),
\end{align*}
where the last equality is due to Stirling's approximation. In order to
evaluate $-n \cdot \log_2(\alpha \! - \! 1)$, we use
Lemma~\ref{lemma:newlow:3:2:alpha:simplification:1} and get
\begin{align*}
  -
  n \cdot \log_2(\alpha \! - \! 1)
    &= -
       n \cdot \log_2 \!
         \Bigg[ \!
           \! \parenv{\that+\Theta\big(n^{-1}\big) \! }
           \cdot
           \parenv{2^{\frac{1}{n-r}}-1} \!
         \Bigg] \\
    &= -
       n 
         \cdot 
         \log_2 \! \parenv{\that+\Theta\big( n^{-1} \big)} \\
    &\quad\ 
       -
       n
         \cdot
         \log_2 \! \parenv{\frac{1}{(1 \! - \! \rho)n\log_2(e)}
                           + O\big( n^{-2} \big)} \\
    &= -
       n
         \cdot
         \log_2(\that)
       +
       n
         \cdot
         \log_2
           \bigl( \log_2(e) \bigr) \\
    &\quad\ 
       +
       n
         \cdot
         \log_2(1 \! - \! \rho)
       +
       n
         \cdot
         \log_2(n)
       +
       o(n),
\end{align*}
where the derivation uses a Taylor series expansion of $2^x$ around
$x = 0$.

Similarly, $(n \! - \! r) \cdot (2r \! - \! n \! + \! 2) \cdot \log_2(\alpha)$
can be rewritten as
\begin{align*}
  &(n \! - \! r) \cdot (2r \! - \! n \! + \! 2) \cdot \log_2(\alpha) \\
  &\quad 
     = (n \! - \! r) \cdot (2r \! - \! n \! + \! 2) \\
  &\quad\quad\ 
     \cdot 
       \log_2 \! 
         \Bigg[
           1
           +
           \parenv{\that+\Theta\big( n^{-1} \big)}
             \cdot 
             \parenv{2^{\frac{1}{n-r}}-1}
         \Bigg]\\
  &\quad 
     = (n \! - \! r) \cdot (2r \! - \! n \! + \! 2) \\
  &\quad\quad\ 
     \cdot 
       \log_2 \!
         \bigg[ 
           1 
           + \parenv{\that+\Theta\big( n^{-1} \big)} \\
       &\quad\quad\quad\quad\quad\quad\quad\quad\quad
             \cdot
             \parenv{\frac{1}{(1 \! - \! \rho)n\log_2(e)} 
                     + 
                     O\big( n^{-2} \big)}
         \bigg] \\
  &\quad 
     = n^2
         \cdot 
         (1 \! - \! \rho)
         \cdot
         (2\rho \! - \! 1)
         \cdot
         \that
         \cdot
         \frac{1}{(1 \! - \! \rho) \cdot n}
         +
         o(n)\\
  &\quad 
     = n
         \cdot 
         (2\rho \! - \! 1)
         \cdot
         \that
       + o(n),
\end{align*}
where the derivation uses a Taylor series expansion of $2^x$ around $x = 0$
and a Taylor series expansion of $\log_2(1+y)$ around $y = 0$.

Combining everything together we get
\begin{align*}
  \log_2\abs{B_{r,n}}
    &\geq
       \log_2\big( \bl_3(\rho,n) \big)
       + 
       o(n),
\end{align*}
as claimed.

\end{document}